\documentclass[preprint]{aastex63}
\usepackage{amsmath}
\shorttitle{EXAMINATION OF POLARIZATION MODE TRANSITIONS}
\shortauthors{McKINNON}

\begin{document}
\title{An Examination of Polarization Mode Transitions in Pulsar Radio Emission}
\author{M. M. McKinnon}
\affiliation{National Radio Astronomy Observatory, Socorro, NM \ 87801 \ \ USA}
\correspondingauthor{M. M. McKinnon}
\email{mmckinno@nrao.edu}

% --------------------------------------------------------------------------------------------

\begin{abstract}

A statistical model is used to determine how stochastic fluctuations in the intensities of 
orthogonal polarization modes contribute to the modulation and depolarization of pulsar 
radio emission. General expressions for the distributions of the Stokes parameters, linear 
polarization, polarization position angle, and fractional polarization are derived when 
the mode intensities follow the same or different probability distributions. The transition 
between modes is examined. When the mode intensities follow the same distribution, the 
fractional linear polarization and modulation index are symmetric about the transition. The 
symmetry is disrupted when the mode intensities follow different distributions. The 
fractional linear polarization is minimum and the mode frequency of occurrence changes 
rapidly at transitions where the mode intensity distributions are the same and the 
modulation index is small. A lower limit on the fractional linear polarization that can be 
attained via the simultaneous occurrence of the modes as a function of modulation index is 
quantified. 

\end{abstract}

% --------------------------------------------------------------------------------------------

\section{INTRODUCTION}

The radio emission from pulsars is highly linearly polarized. The position angle (PA) of
the emission's polarization vector observed in individual pulses and average profiles 
frequently changes discontinuously by $\pi/2$ radians, suggesting the emission is comprised 
of two modes of orthogonal polarization (OPMs; Manchester, Taylor, \& Huguenin 1975). The 
OPMs appear to be a common, if not ubiquitous, feature of the emission (Manchester, Taylor, 
\& Huguenin 1975; Backer \& Rankin 1980; Stinebring et al. 1984). The depolarization of the 
emission with increasing radio frequency (Manchester, Taylor, \& Hueguenin 1973; Morris, 
Graham, \& Sieber 1981) has been attributed to OPMs occurring with comparable frequency and 
similar strength (e.g. Stinebring et al. 1984; McKinnon 1997, 2004; Karastergiou et al. 2002; 
Smits et al. 2006). 

The radio emission is also heavily modulated. The modulation index of the total intensity, 
the ratio of its standard deviation to its mean, $\beta$, can vary over $0.1 < \beta < 2$ 
across a pulsar's pulse and across the pulsar population (e.g. Bartel, Sieber, \& Wolszczan 
1980; Weisberg et al. 1986; Weltevredre et al. 2006, 2007; Burke-Spolaor et al. 2012). The 
modulation is often attributed to subpulse drift, where subpulses move across the overall 
pulse envelope in an organized pattern (e.g. Taylor, Manchester, \& Huguenin 1975; Deshpande 
\& Rankin 2001). In those instances where the observer's sight line tangentially traverses
the pulsar's radio beam, the modulation index resulting from subpulse drift can be relatively 
low and constant across the pulse, where the subpulses can occur regularly from pulse to pulse, 
and increases toward the pulse edges, where the subpulses appear to enter and exit the pulse 
envelope (Rankin 1986; Weisberg et al. 1986). The drift pattern is consistent with a carousel 
of emission sparks circulating about the pulsar polar cap that is ascribed to {\bf E}x{\bf B} 
drift (Ruderman \& Sutherland 1975; Gil \& Sendyk 2000). 

The modulation and OPMs likely conspire to depolarize the emission. With details depending upon 
the polarization of the individual subpulses, the occurrence of OPMs within them, and the subpulse
drift pattern, the aggregate linear polarization will decrease as subpulses within consecutive 
pulses of the star are summed to create the pulsar's average profile. Rankin \& Ramachandran (2003)
proposed that the polarization of PSR B0809+74, an exemplar of drifting pulsars, was determined
by two circulating systems of subpulses, with one system dominated by one polarization mode and 
the other system dominated by the other mode. Their model required the circulating systems to be 
out of phase, with one of the systems shifted further away from the magnetic axis than the other. 
A detailed analysis of the pulsar's polarization by Edwards (2004) confirmed their interpretation, 
but only at 328 MHz. He also found the complex polarization properties of two other drifting pulsars, 
PSR B0320+29 and PSR B0818-13, were difficult to reconcile with the Rankin \& Ramachandran model. 
Clemens \& Rosen (2004, 2008) developed a self-consistent model of subpulse drift and OPMs that 
attributes the drift to nonradial oscillations of the neutron star or its magnetosphere. In their 
model, coherent emission from one of the OPMs, the ``displacement polarization mode", is modulated 
by pulsational displacements. The emission from the other OPM, the ``velocity polarization mode", 
is modulated by pulsational velocities. Their model produces a reasonable facsimile of the drifting 
subpulses and polarization in PSR B0940+10 (Rosen \& Clemens 2008) and PSR B0809+74 (Rosen \& 
Demorest 2011). An important clarification to be made here is the modulation and depolarization of 
the emission by subpulse drift, regardless of its origin, is a {\it systematic} process.

Not all pulsars, however, exhibit subpulse drift (Weltevrede et al. 2006, 2007), and the switching
between OPMs is generally regarded as a {\it stochastic} process (Cordes, Rankin, \& Backer 1978). 
The emission has been accurately modeled as a stochastic sequence of polarization states with 
structure having several characteristic timescales (Cordes 1976a, 1976b; Cordes \& Hankins 1977). 
Additionally, while some assessments of subpulse drift assume the radiation from a subpulse is 
steady and unmodulated (e.g. Jenet \& Gil 2003), the observed subpulse structure is variable and 
complex and cannot be attributed  to instrumental noise (see, e.g., Figures 2-5 in Gil \& Sendyk 
(2000) for PSRs B0943+10, B2303+30, B2319+60, and B0031-07; Figure 1 of Cordes (1976a) for PSR 
B2016+28; and the discussion of a stochastic component in the emission of PSR B0943+10 in Rosen 
\& Clemens (2008)). Therefore, while the process that modulates and depolarizes the emission 
from some pulsars may be systematic, it may be stochastic, or both, in others.

A statistical model for the polarization of pulsar radio emission established a connection 
between OPMs and intensity modulation (McKinnon \& Stinebring 1998, hereafter MS). The model 
assumes the emission is comprised of two independent, simultaneously occurring, completely 
polarized orthogonal modes. The model treats the mode intensities as random variables (RVs) to 
account for the emission's variability, and consequently focuses on stochastic fluctuations of 
the emission. The details of the model, its theoretical underpinnings, and the recipe for its 
implementation can be found in MS and McKinnon (2022; hereafter M22). Since the modes are 
orthogonal, their polarization vectors are antiparallel in the Poincar\'e sphere; therefore, 
the model predicts that polarization fluctuations occur along a diagonal in the sphere. One 
might reasonably conclude that polarization fluctuations are large when the total intensity is 
heavily modulated, but absent the model, it is not intuitively obvious that the polarization 
fluctuations have a preferred orientation in the sphere, as the model requires. When multiple 
samples of the Stokes parameters Q, U, and V recorded at a specific pulse phase are plotted in 
the Poincar\'e sphere, the shape of the resulting cluster of data points is a prolate ellipsoid. 
Absent any other emission component, the size of the ellipsoid's major axis is determined by a 
combination of the polarization fluctuations and instrumental noise, while the sizes of the 
ellipsoid's minor axes are set by the instrumental noise alone (McKinnon 2004, hereafter M04). 
The axial ratio of the major axis is approximately $(1+s^2\beta^2)^{1/2}$, where $s$ is the 
signal-to-noise ratio of the total intensity (M04) and $\beta$ is its modulation index. The 
axial ratio is large where the signal-to-noise ratio and modulation index are also large. 
This prediction was found to be generally true for PSR B2020+28 and PSR B1929+10 (M04) and 
in the outriders of PSR B0329+54 (Edwards \& Stappers 2004, hereafter ES04). The model and 
observations suggest randomly fluctuating OPMs play a role in determining both the observed 
polarization and modulation of the emission.

The MS statistical model can be used to derive distributions of the Stokes parameters of 
the combined radiation, its linear polarization, fractional polarization, and PA given the 
distributions of the individual mode intensities. In the implementation of the model in MS and 
M22, the statistical character of the mode intensities was assumed to be the same. The mode 
intensities were assumed to be Gaussian RVs with identical standard deviations in MS. The 
distributions produced by the model were varied by altering the mean intensity of one mode 
relative to that of the other. In M22, the mode intensities were assumed to be exponential RVs 
to account for the heavy modulation of the total intensity, the correlation of the Stokes 
parameters, and observed asymmetries in distributions of total intensity, polarization, and 
fractional polarization. The primary conclusion from M22 was the mode intensities must have 
different variances to explain what is observed. The difference in variances was attributed to 
either different emission mechanisms for the modes or to mode-dependent propagation or scattering 
effects in the pulsar magnetosphere. Summarizing, previous implementations of the model explored 
the effects of varying the means (MS) and variances (M22) of mode intensities that followed the 
same distribution. The next logical step in this line of investigation is to explore how mode 
intensities of different statistical character might affect what is observed, as there appears 
to be no a priori reason for them to follow the same distribution. 

The primary objectives of this paper are twofold. The first is to explore the polarization and 
modulation properties of the emission when the mode intensities follow different probability 
distributions. The second is to conduct a detailed examination of a transition between modes 
in an average pulse profile.  The analysis assumes the functional form of a mode's intensity 
distribution is retained throughout the short duration of a transition, but the relative scaling 
of the two distributions varies across it. The variation in the relative scaling of the two 
distributions drives the transition between the modes and causes the modulation index to vary 
across the transition.

The paper is organized as follows. The model is implemented in Section~\ref{sec:implement}
assuming the mode intensities follow different Erlang distributions. The Erlang distribution was 
chosen for the analysis because it can produce a range of modulation indices that approaches 
what is observed in pulsar radio emission, and it allows analytical solutions within the context 
of the statistical model. The assumption of Erlang mode intensities also enables a general
implementation of the MS statistical model of which the Gaussian and exponential intensities 
considered in MS and M22 are specific cases. Distributions of the emission's Stokes parameters, 
linear polarization, fractional polarization, and polarization position angle at a given pulse 
phase are derived. Statistical parameters that characterize the emission and its polarization are 
also derived. The salient features of the analysis are highlighted in Section~\ref{sec:features}. 
A detailed examination is made of a mode transition, and the effectiveness of randomly fluctuating 
OPMs in depolarizing the emission is evaluated. The statistical parameters and distributions of 
fractional polarization derived from Erlang distributions are compared with their counterparts 
derived from Gaussian distributions. The results of the analysis are discussed and compared with 
observations in Section~\ref{sec:discuss}. Conclusions are summarized in Section~\ref{sec:conclude}. 

% --------------------------------------------------------------------------------------------

\section{MODEL IMPLEMENTATION}
\label{sec:implement}

\subsection{Review of the Erlang Distribution}

The Erlang distribution arises from the sum of $n$ independent, identically distributed, 
exponential RVs (e.g. Ross 1984, p. 171), and is given by

\begin{equation}
f(x,\mu,n) = \frac{x^{n-1}}{\Gamma(n)\mu^n}\exp{\left(-\frac{x}{\mu}\right)}, \qquad x\ge 0,
\end{equation}

\noindent where $\mu$ is a positive scaling factor, $n$ is a positive integer representing 
the order of the distribution, and $\Gamma(n)=(n-1)!$ is the gamma function. The Erlang 
distribution is the exponential distribution when $n=1$. From the central limit theorem, 
the distribution approaches the Gaussian distribution when $n$ becomes very large. The 
$k^{th}$ moment of the distribution is

\begin{equation}
\langle x^k\rangle = \frac{\Gamma(n+k)}{\Gamma(n)}\mu^k.
\label{eqn:moment}
\end{equation}

\noindent The mean of the distribution is $n\mu$, and its variance is $\sigma^2=n\mu^2$. 
The modulation index of the Erlang distribution is $\beta=1/\sqrt{n}$, which is independent 
of $\mu$ and decreases with increasing $n$. 

Since the distribution of the sum of independent RVs is the convolution of their individual
distributions (e.g. Papoulis 1965, p. 189), the distribution of the sum of two Erlang RVs 
having orders $n_a$ and $n_b$ and identical scaling factors, $\mu$, is also an Erlang 
distribution with a scaling factor $\mu$ and order $n=n_a+n_b$.

\begin{equation}
f(x,\mu,n_a)*f(x,\mu,n_b) = f(x,\mu,n_a+n_b)
\label{eqn:equal_mu}
\end{equation}

% --------------------------------------------------------------------------------------------

\subsection{Definitions and Nomenclature}
\label{sec:nomenclature}

The implementation of the statistical model in MS and M22 focused on one of the orthogonal 
modes remaining the primary (dominant) mode. To be specific, the primary mode as discussed 
in this paper is defined as the polarization mode that occurs most frequently, as in a 
PA histogram. By not exploring the possibility of the other mode becoming the primary, 
MS and M22 did not exploit the full capability of the model. To do so requires minor 
revisions to the definitions and nomenclature used in the analysis. These revisions do not 
alter the results or conclusions described in MS and M22. Here, the modes are designated 
as A and B, without the presumption that one is the primary or secondary mode. When the 
OPMs are completely linearly polarized, the instantaneous total intensity is the sum of the 
RVs representing the mode intensities, $I=X_A+X_B$, and the Stokes parameter Q is their 
difference, $Q=X_A-X_B$ (MS). The remaining Stokes parameters, U and V, are temporarily 
assumed to be fixed at zero to simplify the analysis. So defined, the Stokes parameters I 
and Q will generally, but not always, be correlated when mode A is the primary mode, and 
will be anticorrelated when mode B is the primary (see Sections~\ref{sec:parms} 
and~\ref{sec:transition}). The scaling factor and order of the mode A Erlang distribution 
are designated as $a$ and $n_a$, respectively. Similarly, the scaling factor and order of 
mode B are $b$ and $n_b$.

MS and M22 defined a parameter $M$ as the ratio of the mode mean intensities, 
$M=\langle X_A\rangle/\langle X_B\rangle$, where the angular brackets denote an 
average over pulse number at a given pulse phase. That definition is retained in this 
analysis. The range of $M$ in MS and M22 was restricted to $1\le M\le\infty$, because only 
one of the modes was viewed as the primary in those analyses. Here, the range of $M$ is 
revised to $0\le M\le\infty$ to accommodate the possibility that either mode can be the 
primary. M22 also defined a parameter $m$ as the ratio of the mean Stokes parameter 
Q to the mean total intensity.

\begin{equation}
m = \frac{\mu_{\rm Q}}{\mu_{\rm I}}
  =  \frac{\langle X_A\rangle-\langle X_B\rangle}{\langle X_A\rangle+\langle X_B\rangle}
  = \frac{M-1}{M+1}
\end{equation}

\noindent Its range must be revised from $0\le m\le 1$ in M22 to $-1\le m\le 1$. The 
emission is comprised solely of mode A when $m=1$ ($M=\infty$), and is comprised solely of 
mode B when $m=-1$ ($M=0$). The equations derived in the analysis are simplified when a 
parameter $C$ is defined as the ratio of the distribution scaling factors, $C=a/b$, along 
with its complementary parameter $c=(C-1)/(C+1)$. When $n_a=n_b$, $M=C$ and $m=c$.

Instrumental noise is not included in this particular implementation of the model. The 
effects of instrumental noise and its treatment in the analysis are discussed in MS, 
McKinnon (2002), M04, and M22.

% --------------------------------------------------------------------------------------------

\subsection{Distributions of the Stokes Parameters, Linear Polarization, and Fractional 
Polarization}
\label{sec:dist}

The MS statistical model assumes the mode intensities are statistically independent. Therefore, 
the distribution of the total intensity, $f_{\rm I}$, at a given pulse phase is the convolution 
of the mode intensity distributions. The distribution resulting from the convolution of two 
Erlang distributions was derived by Kadri \& Smaili (2015). Their Equation 17 rewritten with 
the nomenclature used in this paper is 

\begin{eqnarray}
f_{\rm I}(y,n_a,n_b) & = & \int_0^y f(y-x,a,n_a)f(x,b,n_b)dx \nonumber \\
      & = & (-1)^{n_a}\left(\frac{b}{a-b}\right)^{n_a}\left(\frac{a}{a-b}\right)^{n_b}
      \Biggl[\sum_{k=1}^{n_a}\binom{n_a+n_b-k-1}{n_b-1}\left(\frac{b-a}{b}\right)^k f(y,a,k)
      \nonumber \\
 & + & \sum_{k=1}^{n_b}\binom{n_a+n_b-k-1}{n_a-1}\left(\frac{a-b}{a}\right)^k f(y,b,k)\Biggr],
                     \qquad y\ge 0,
\label{eqn:fI}
\end{eqnarray}

\noindent where the leading parenthetical term in each summation is the binomial coefficient.
Equation~\ref{eqn:fI} shows that $f_{\rm I}$ is an additive combination of weighted Erlang 
distributions, with the orders of the distributions ranging from $n=1$ to the larger of $n_a$ 
or $n_b$. The equation is valid as long as $a\ne b$. When $a=b$, $f_{\rm I}$ is a single
Erlang distribution with $n=n_a+n_b$, as stipulated by Equation~\ref{eqn:equal_mu}.

The distribution of the Stokes parameter Q, $f_{\rm Q}$, is the correlation of the mode 
intensity distributions (MS). The distribution can be derived by replacing the term $(y+x)^n$ 
in the correlation with the binomial expansion.

\begin{eqnarray}
f_{\rm Q}(y,n_a,n_b) & = & \int_{-\infty}^\infty f(y+x,a,n_a)f(x,b,n_b)dx \nonumber \\
         & = & \left(\frac{b}{a+b}\right)^{n_a-1}\left(\frac{a}{a+b}\right)^{n_b}
         \sum_{k=0}^{n_a-1}\binom{n_a+n_b-k-2}{n_b-1}\left(\frac{a+b}{b}\right)^k f(y,a,k+1),
                       \quad y\ge 0, \nonumber \\
             &   & \left(\frac{a}{a+b}\right)^{n_b-1}\left(\frac{b}{a+b}\right)^{n_a}
  \sum_{k=0}^{n_b-1}\binom{n_a+n_b-k-2}{n_a-1}\left(\frac{a+b}{a}\right)^k f(\lvert y\rvert,b,k+1),
                       \quad y<0
\label{eqn:fQ}
\end{eqnarray}

\noindent Equation~\ref{eqn:fQ} shows that $f_{\rm Q}$ is also an additive combination of 
weighted Erlang distributions. 

As noted in Section~\ref{sec:nomenclature}, the MS statistical model also assumes the 
Stokes parameter U is equal to zero, so that the linear polarization is equal to the absolute 
value of the Stokes parameter Q, $L=(Q^2+U^2)^{1/2}=|Q|$. The distribution of linear 
polarization, $f_{\rm L}$, is the distribution of Stokes Q folded about $y=0$. Therefore, 
it is the sum of the two terms in Equation~\ref{eqn:fQ} for positive values of $y$. 

\begin{eqnarray}
f_{\rm L}(y,n_a,n_b) & = & \left(\frac{b}{a+b}\right)^{n_a-1}\left(\frac{a}{a+b}\right)^{n_b}
         \sum_{k=0}^{n_a-1}\binom{n_a+n_b-k-2}{n_b-1}\left(\frac{a+b}{b}\right)^k f(y,a,k+1)
         \nonumber \\
         & + & \left(\frac{a}{a+b}\right)^{n_b-1}\left(\frac{b}{a+b}\right)^{n_a}
         \sum_{k=0}^{n_b-1}\binom{n_a+n_b-k-2}{n_a-1}\left(\frac{a+b}{a}\right)^k f(y,b,k+1),
                       \quad y\ge 0
\label{eqn:fL}
\end{eqnarray}

When the orthogonal modes are elliptically polarized, the instantaneous values of the 
Stokes parameters Q, U, and V at a given pulse phase are simply scaled versions of $X_A-X_B$ 
(M22)

\begin{equation}
Q=\cos(2\psi_o)\cos(2\chi_o)(X_A-X_B)
\end{equation}

\begin{equation}
U=\sin(2\psi_o)\cos(2\chi_o)(X_A-X_B)
\end{equation}

\begin{equation}
V=\sin(2\chi_o)(X_A-X_B),
\label{eqn:V}
\end{equation}

\noindent where $\psi_o$ is the PA of the mode A polarization vector and $\chi_o$ is its 
ellipticity angle\footnote{In M22, the colatitude, $\theta$, of the mode A polarization 
vector in the Poincar\'e sphere was used to represent the polarization's ellipticity. The 
conventional ellipticity angle, $\chi$, is used to represent ellipticity in this 
analysis.} (EA; $\chi_o=0$ when the modes are linearly polarized, and $\chi_o=\pm\pi/4$ 
when the modes are circularly polarized). The distributions of Q, U, and V are then scaled 
versions of $f_{\rm Q}$ given by Equation~\ref{eqn:fQ}. The Stokes parameter I remains the 
sum of $X_A$ and $X_B$. Consequently, the distributions of all four Stokes parameters and 
the linear polarization are additive combinations of weighted Erlang distributions. 

The procedure for deriving the distributions of fractional polarization is outlined in M22. 
Without loss of generality, the PA can be set to zero to simplify the analysis. The distribution 
of fractional circular polarization, $f_{\rm mv}$, is derived from the joint probability density 
of the Stokes parameters I and V. The distribution of the fractional linear polarization, 
$f_{\rm ml}$, is derived from the joint probability density of the Stokes parameters I and Q. 
The distributions follow the general forms of 

\begin{equation}
f_{\rm mv}(z,c,n_a,n_b,\chi_o)  = K(c,n_a,n_b)\sin(2\chi_o)
\left[\frac{(\sin(2\chi_o)+z)^{n_a-1}(\sin(2\chi_0)-z)^{n_b-1}}{(\sin(2\chi_o)-cz)^{n_a+n_b}}\right],
\label{eqn:fmv}
\end{equation}

\begin{eqnarray}
f_{\rm ml}(z,c,n_a,n_b,\chi_o) & = & K(c,n_a,n_b)\cos(2\chi_o)
      \Biggl[{(\cos(2\chi_o)+z)^{n_a-1}(\cos(2\chi_o)-z)^{n_b-1}\over{(\cos(2\chi_o)-cz)^{n_a+n_b}}}
      \nonumber \\ 
& + & {(\cos(2\chi_o)+z)^{n_b-1}(\cos(2\chi_o)-z)^{n_a-1}\over{(\cos(2\chi_o)+cz)^{n_a+n_b}}}\Biggr],
\label{eqn:fml}
\end{eqnarray}
               
\noindent where $K(c,n_a,n_b)$ is a normalization constant, given by
  
\begin{equation} 
K(c,n_a,n_b) = 2\left(\frac{1-c}{2}\right)^{n_a}\left(\frac{1+c}{2}\right)^{n_b}
                \frac{(n_a+n_b-1)!}{(n_a-1)!(n_b-1)!}.
\end{equation}

\noindent The distributions of fractional polarization are functions of four free parameters: 
$c$, $n_a$, $n_b$, and $\chi_o$. The range of the independent variable in $f_{\rm mv}$ is 
$-\sin(2\chi_o)\le z\le\sin(2\chi_o)$, and the range of the independent variable in $f_{\rm ml}$ 
is $0\le z\le\cos(2\chi_o)$ (M22). When the orders of the mode intensity distributions are the 
same ($n_a=n_b=n$), $m=c$, and the fractional polarization distributions are

\begin{equation}
f_{\rm mv}(z,m,n,n,\pi/4) = \frac{(1-m^2)^n}{2^{2n-1}}\frac{(2n-1)!}{(n-1)!^2}
                            \frac{(1-z^2)^{n-1}}{(1-mz)^{2n}},
\end{equation}

\begin{equation}
f_{\rm ml}(z,m,n,n,0) = \frac{(1-m^2)^n}{2^{2n-1}}\frac{(2n-1)!}{(n-1)!^2}
                        \left[\frac{(1+mz)^{2n} + (1-mz)^{2n}}{(1-m^2z^2)^{2n}}\right]
                         (1-z^2)^{n-1}.
\end{equation}

% --------------------------------------------------------------------------------------------

\subsection{Model Statistical Parameters}
\label{sec:parms}

Parameters can be derived to quantify the statistical properties of the emission and its 
polarization (M22). For OPMs that are completely linearly polarized, the parameters include 
the mean of Stokes Q relative to the mean total intensity, the modulation index of the total 
intensity, the correlation coefficient of the Stokes parameters I and Q, the frequency of 
occurrence of each mode, and the mean fractional linear polarization. The mean of the Stokes 
parameter Q, $\mu_{\rm Q}$, normalized by the mean total intensity, $\mu_{\rm I}$, is

\begin{equation}
\bar{\rm Q}(c,n_a,n_b) = m = \frac{\mu_{\rm Q}}{\mu_{\rm I}} 
                       = \frac{n_a(1+c)-n_b(1-c)}{n_a(1+c)+n_b(1-c)}.
\label{eqn:Qbar}
\end{equation}

\noindent The normalized Stokes parameter Q is always equal to the parameter $m$. It ranges 
from $\bar{\rm Q}=1$ at $c=1$ to $\bar{\rm Q}=-1$ at $c=-1$. The transition between modes 
occurs at $\bar{\rm Q}=0$ where the mode mean intensities are equal.

The modulation index of the total intensity is its standard deviation relative to its mean.

\begin{equation}
\beta(c,n_a,n_b) = \frac{\sigma_{\rm I}}{\mu_{\rm I}} 
                 = \frac{[n_a(1+c)^2+n_b(1-c)^2]^{1/2}}{n_a(1+c)+n_b(1-c)}
\label{eqn:beta}
\end{equation}

\noindent The modulation index ranges from that of mode A ($\beta=1/\sqrt{n_a}$) at $c=1$
to that of mode B ($\beta=1/\sqrt{n_b}$) at $c=-1$. The minimum value of $\beta$ always 
occurs at $c=0$ (i.e. when $a=b$), where it is given by $\beta=1/\sqrt{n_a+n_b}$. The
distribution of the total intensity at $c=0$ is a single Erlang distribution (see 
Equation~\ref{eqn:equal_mu}).

The Stokes parameters I and Q are generally covariant. The I-Q correlation coefficient is 

\begin{equation}
r_{\rm IQ}(c,n_a,n_b) = \frac{\sigma_a^2-\sigma_b^2}{\sigma_a^2+\sigma_b^2}
                      = \frac{n_a(1+c)^2-n_b(1-c)^2}{n_a(1+c)^2+n_b(1-c)^2}.
\label{eqn:rIQ}
\end{equation}

\noindent The correlation coefficient ranges from $r_{\rm IQ}=1$ at $c=1$ to 
$r_{\rm IQ}=-1$ at $c=-1$. It is equal to zero when the variances of the mode intensities
are equal.

When $n_a=n_b=n$, the parameter $m$ is equal to $c$, and $r_{\rm IQ}$ and $\beta$ are 
given by

\begin{equation}
r_{\rm IQ}(m)=\frac{2m}{1+m^2},
\label{eqn:Corr}
\end{equation}

\begin{equation}
\beta(m,n)=\left(\frac{1+m^2}{2n}\right)^{1/2}.
\end{equation}

\noindent The correlation coefficient and normalized mean of Stokes Q are independent of 
$n$. The modulation index remains a function of $n$, and varies by a factor of $\sqrt{2}$ 
over the full range of $m$ for a given value of $n$.

The distribution of polarization position angle consists of two delta functions separated by 
$\pi/2$ radians. The amplitude of the delta function associated with mode A, or the frequency 
of occurrence of mode A, is the probability that the Stokes parameter Q exceeds zero (MS).

\begin{eqnarray}
\nu_a(c,n_a,n_b) & = & \int_0^\infty f_{\rm Q}(y,n_a,n_b)dy \nonumber \\
                 & = & \left(\frac{1-c}{2}\right)^{n_a-1}\left(\frac{1+c}{2}\right)^{n_b}
                 \sum_{k=0}^{n_a-1}\binom{n_a+n_b-k-2}{n_b-1}\left(\frac{2}{1-c}\right)^k
\label{eqn:nuA}
\end{eqnarray}

\noindent The integration in Equation~\ref{eqn:nuA} is straightforward, because the integral of
$f(y,a,k+1)$ for each index, $k$, in the summation contained within $f_{\rm Q}$ is equal to one. 
The frequency of occurence of mode B is $\nu_b=1-\nu_a$, and the difference in frequency of 
occurrence is $\Delta\nu=\nu_a-\nu_b=2\nu_a-1$. The equation for the difference in mode 
frequency of occurrence has a simple analytical form when either of $n_a$ or $n_b$ is equal to 
one. When $n_a=1$, $\Delta\nu$ is

\begin{equation}
\Delta\nu(c,1,n_b) = 2\left(\frac{1+c}{2}\right)^{n_b} - 1.
\label{eqn:deltanu_b}
\end{equation}

\noindent When $n_b=1$, $\Delta\nu$ is

\begin{equation}
\Delta\nu(c,n_a,1) = 1 - 2\left(\frac{1-c}{2}\right)^{n_a}.
\label{eqn:deltanu_a}
\end{equation}
 
The mean linear polarization, $\mu_{\rm L}$, normalized by the mean total intensity, or the mean 
fractional linear polarization, can be calculated using the expression for the first moment of the 
Erlang distribution given by Equation~\ref{eqn:moment}.

\begin{eqnarray}
\bar{\rm L}(c,n_a,n_b) = \frac{\mu_{\rm L}}{\mu_{\rm I}}
           & = & \frac{2}{n_a(1+c)+n_b(1-c)}\left(\frac{1-c}{2}\right)^{n_a}
                 \left(\frac{1+c}{2}\right)^{n_b} \nonumber \\
  & \times & \Biggl[\left(\frac{1+c}{1-c}\right)\sum_{k=0}^{n_a-1}\binom{n_a+n_b-k-2}{n_b-1} 
                   \left(\frac{2}{1-c}\right)^k(k+1) \nonumber \\
        & + & \left(\frac{1-c}{1+c}\right)\sum_{k=0}^{n_b-1}\binom{n_a+n_b-k-2}{n_a-1} 
              \left(\frac{2}{1+c}\right)^k(k+1)\Biggr] 
\label{eqn:Lbar}
\end{eqnarray}

\noindent The equation for $\bar{\rm L}$ also has a simple analytical form when either 
of $n_a$ or $n_b$ is equal to one.

\begin{equation}
\bar{\rm L}(c,1,n_b) = \frac{4}{n_b(1-c)+(1+c)}\left(\frac{1+c}{2}\right)^{n_b+1} 
                     - \bar{\rm Q}(c,1,n_b),
\label{eqn:LbarB}
\end{equation}

\begin{equation}
\bar{\rm L}(c,n_a,1) = \frac{4}{n_a(1+c)+(1-c)}\left(\frac{1-c}{2}\right)^{n_a+1} 
                     + \bar{\rm Q}(c,n_a,1).
\label{eqn:LbarA}
\end{equation}

\noindent All of the statistical parameters are functions only of $c$, $n_a$, and $n_b$. 
They are independent of the values of $a$ and $b$. 

\begin{figure}
\plotone{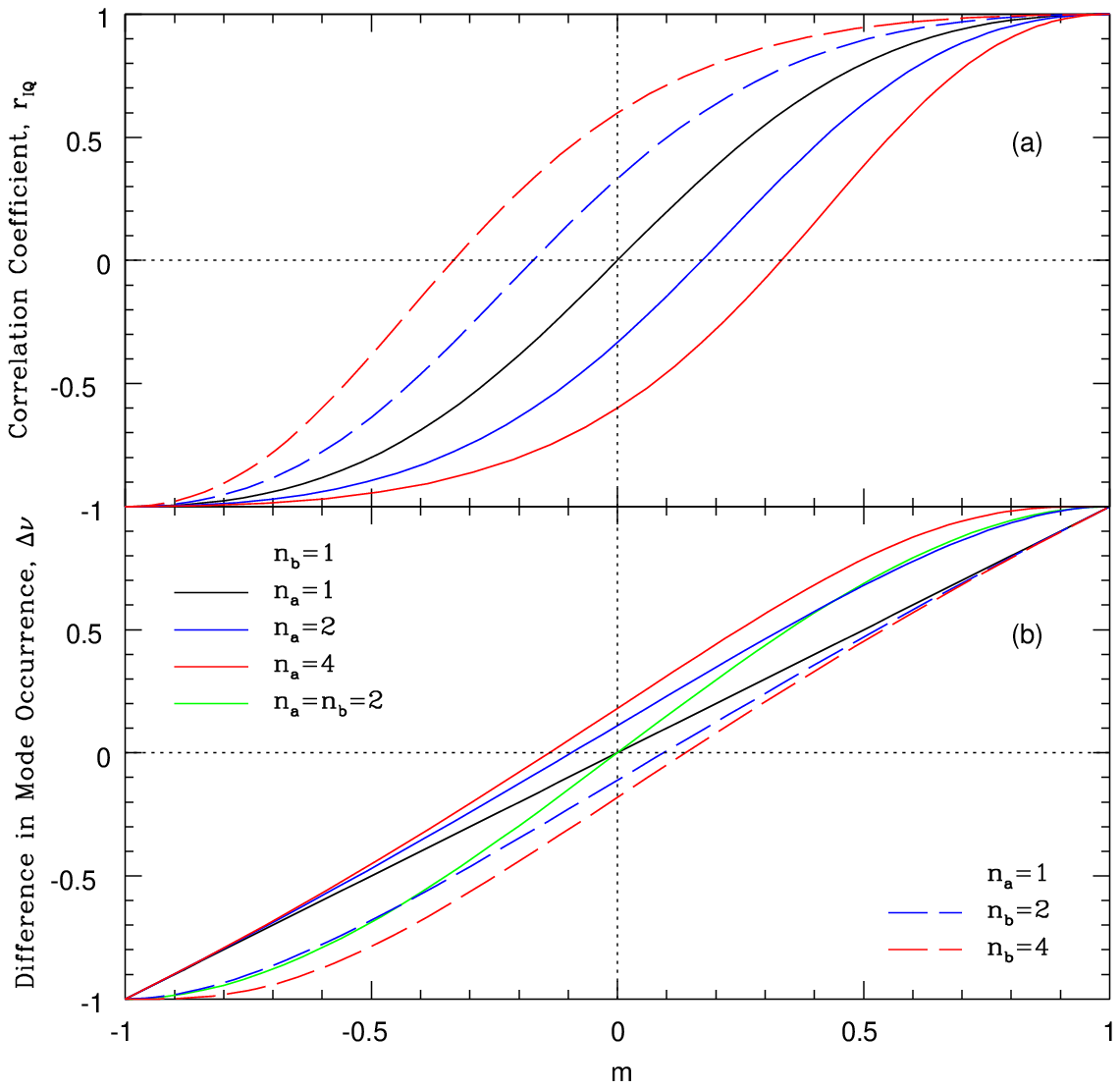}
\caption{The behavior of the I-Q correlation coefficient (a) and difference in mode frequency 
of occurrence (b) across a mode transition. The solid black, blue, and red lines are drawn for 
$n_b=1$, with $n_a$ taking on values of $n_a=1,2,4$, respectively. The dashed lines are drawn 
for $n_a=1$, with $n_b$ taking on values of $n_b=2,4$. The solid green line in (b) shows 
$\Delta\nu$ when $n_a=n_b=2$. The dependence of $r_{\rm IQ}$ upon $m$ for $n_a=n_b=2$ is 
identical to that for $n_a=n_b=1$ (see Equation~\ref{eqn:Corr}).}
\label{fig:Qrnu}
\end{figure}

\begin{figure}
\plotone{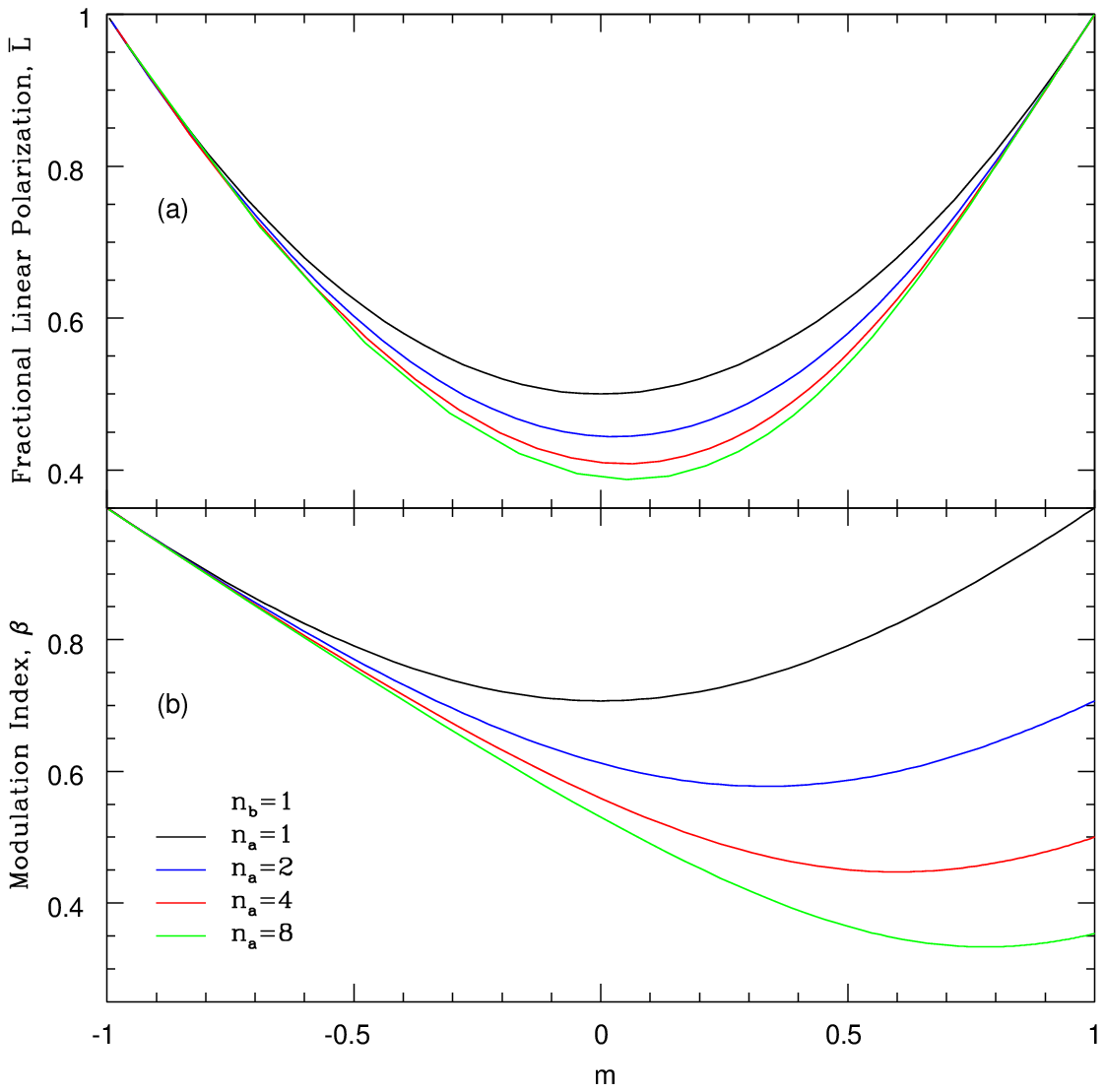}
\caption{Change in fractional linear polarization (a) and modulation index (b) across a mode 
transition. The curves are drawn for $n_b=1$ with $n_a$ taking on values of $n_a=1,2,4,$ and $8$.}
\label{fig:LinBeta}
\end{figure}

% -----------------------------------------------------------------------------------------

\subsection{Variations in the Statistical Parameters across a Mode Transition}

The statistical parameters described in Section~\ref{sec:parms} are average properties
calculated at a given pulse phase. They were determined, in part, from the distributions of 
the Stokes parameters and linear polarization derived in Section~\ref{sec:dist}. The 
variations in the statistical parameters across a mode transition within an average profile 
are illustrated in this section.

The ratio of the mode mean intensities, $M$, changes with variations in the relative scaling
of the mode intensity distributions. Changes in $M$, or alternatively the parameter $m$, with 
pulse phase drive the transition between the modes. Since $m$ must change sufficiently to 
complete the transition, it may be viewed as a proxy for pulse phase over the duration of the 
transition. The behavior of the statistical parameters across the overall mode transition can 
be simulated and compared by plotting them as functions of $m$. The transition between modes
at $m=0$ serves as the fiducial point for the comparison.

Figure~\ref{fig:Qrnu} shows the behavior of $\Delta\nu$ and $r_{\rm IQ}$ across a mode 
transition for different values of $n_a$ and $n_b$. Panel (a) of the figure shows 
$r_{\rm IQ}$ calculated from Equation~\ref{eqn:rIQ}. Panel (b) shows $\Delta\nu$ as 
calculated from Equations~\ref{eqn:deltanu_b} and~\ref{eqn:deltanu_a}. A transition between
modes occurs at $m=0$ in both panels of the figure. When $n_a=n_b$, $\Delta\nu$ and 
$r_{\rm IQ}$ are antisymmetric about $m$, where they are also equal to zero. The antisymmetry 
is disrupted when $n_a\ne n_b$. The parameters in the figure generally pass through zero at 
different values of $m$. In panel (b), the change in primary mode, as denoted by 
$\Delta\nu=0$, generally does not coincide with the mode transition. Mode A is the primary 
mode for values of $m$ where $\Delta\nu >0$, and mode B is the primary mode where 
$\Delta\nu <0$. Panel (b) also shows a mode can be the primary over a larger range of $m$ 
when $n_a\ne n_b$ than when $n_a=n_b$. Contrary to what one might initially surmise from 
the model's definitions of I, Q, and $r_{\rm IQ}$, the figure shows there are values of 
$m$ where mode A (B) is the primary, although the Stokes parameters I and Q are 
anticorrelated (correlated). 

Figure~\ref{fig:LinBeta} shows the change in $\bar{\rm L}$ and $\beta$ across a mode 
transition for $n_b=1$ and $n_a=1,2,4,$ and $8$. The curves for $\bar{\rm L}$ shown in panel 
(a) of the figure were produced from Equation~\ref{eqn:LbarA}. The curves for $\beta$ 
in panel (b) were produced using Equation~\ref{eqn:beta}. When $n_a=n_b$, $\bar{\rm L}$ 
and $\beta$ are symmetric about $m=0$ and pass through minima there. The symmetry of 
$\bar{\rm L}$ and $\beta$ about $m=0$ is disrupted when $n_a\ne n_b$. Panel (a) shows 
that the minimum of fractional linear polarization decreases as the sum of $n_a$ and 
$n_b$ increases. The location of the minimum moves to increasing values of $m$ as $n_a$ 
increases. Panel (b) shows that the range of possible values of $\beta$ is greatest when 
the order of one mode's distribution is $n=1$ and the order of the other distribution is 
large. 

The symmetry properties of the statistical parameters at a mode transition when
$n_a=n_b$ and $n_a\ne n_b$ are summarized in Table 1.

\begin{deluxetable}{ccc}
\tablenum{1}
\tablecaption{Symmetry Properties of Statistical Parameters across a Mode Transition}
\tablehead{\colhead{Mode Erlang Order} & \colhead{$n_a=n_b$} & \colhead{$n_a\ne n_b$}}
\startdata
Normalized Stokes Q, $m=\bar{\rm Q}$  & Antisymmetric & Antisymmetric \\
Difference in mode occurrence, $\Delta\nu$ & Antisymmetric & Asymmetric \\
I-Q correlation coefficient, $r_{\rm IQ}$ & Antisymmetric & Asymmetric \\
Normalized linear polarization, $\bar{\rm L}$ & Symmetric & Asymmetric \\
Modulation index, $\beta$ & Symmetric & Asymmetric \\
\hline
Mean intensity ratio, $M$  & $M=C$ & $M\ne C$ \\
Normalized Stokes Q, $m$  & $m=c$ & $m\ne c$ \\
\enddata
\end{deluxetable}

% --------------------------------------------------------------------------------------------

\section{ANALYSIS FEATURES}
\label{sec:features}

\subsection{Detailed Examination of a Mode Transition}
\label{sec:transition}

\begin{figure}
\plotone{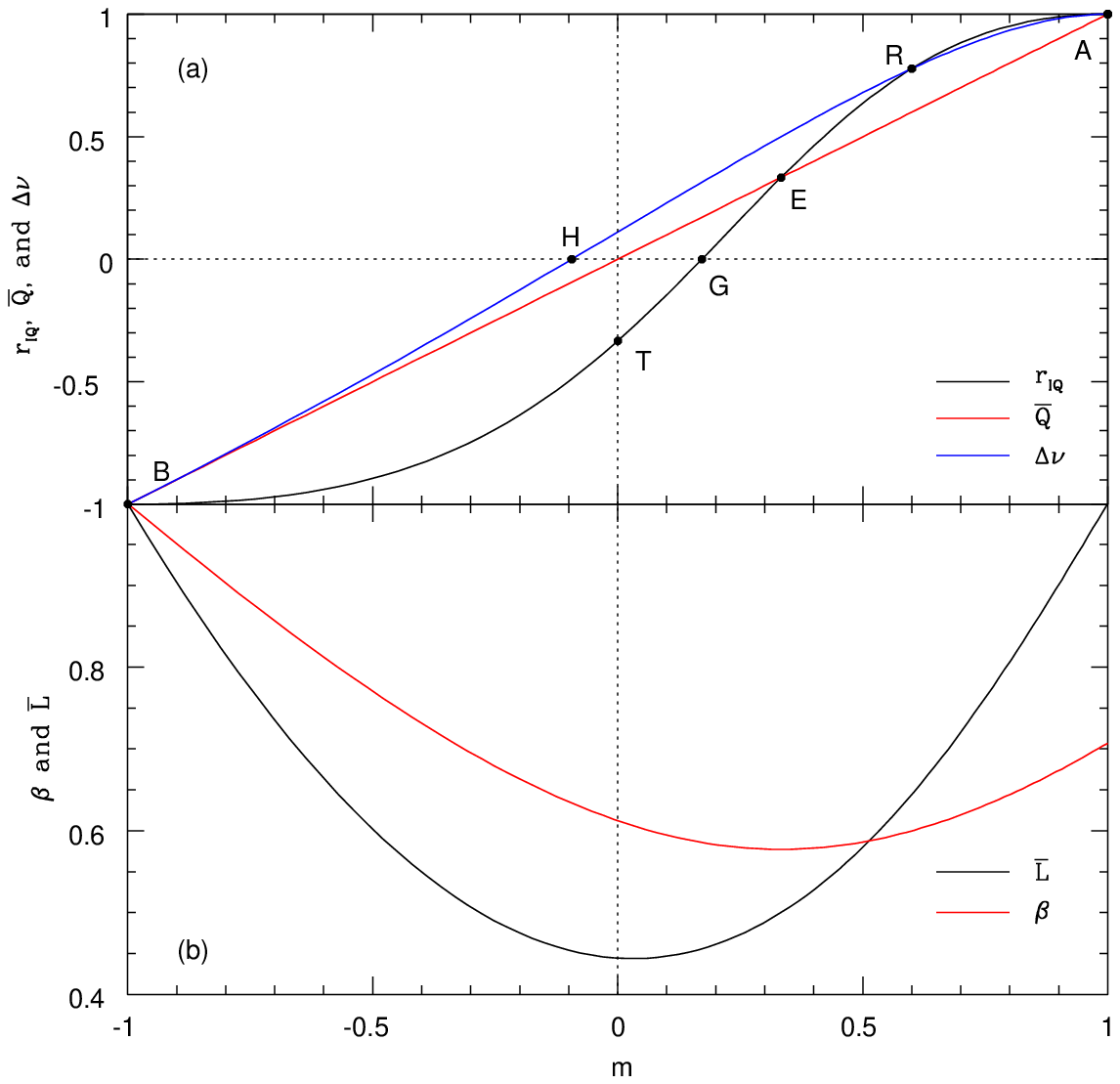}
\caption{Variations in the statistical parameters across a mode transition when $n_a=2$ 
and $n_b=1$. Panel (a) shows the change in I-Q correlation coefficient ($r_{\rm IQ}$), 
normalized Stokes parameter Q ($\bar{\rm Q}$), and difference in mode frequency of occurrence 
($\Delta\nu$) with the parameter $m$. Panel (b) shows the dependence of the normalized linear 
polarization ($\bar{\rm L}$) and modulation index ($\beta$) upon $m$. Decreasing the value of 
$m$ drives the transition from mode A to mode B. See the text and Table 2 for explanations of 
the points labeled in the figure.}
\label{fig:hysteresis}
\end{figure}

The distributions and statistical parameters derived in Sections~\ref{sec:dist} 
and~\ref{sec:parms} allow a detailed examination of a mode transition when the mode 
intensities follow different Erlang distributions. A transition from mode A to mode B is 
illustrated in Figure~\ref{fig:hysteresis} when $n_a=2$ and $n_b=1$. Panel (a) of the figure 
shows the I-Q correlation coefficient, normalized mean of the Stokes parameter Q, and the 
difference in mode frequency of occurrence as functions of $m$. Panel (b) shows the modulation 
index and normalized mean of the linear polarization. At point A in panel (a) of the figure, 
$m=c=1$ and the emission is comprised solely of mode A. The emission is completely polarized,
and the Stokes parameters I and Q are completely correlated. The modulation index is equal to 
that of mode A ($\beta=1/\sqrt{n_a}$). The statistical parameters decrease as $m$, or $c$, 
decreases. At point R in the figure, $C=n_a/n_b$, and $c$ is

\begin{equation}
c = \frac{n_a-n_b}{n_a+n_b}.
\end{equation}

\noindent The values of $m$ and $r_{\rm IQ}$ at point R are always 

\begin{equation}
m = \frac{n_a^2-n_b^2}{n_a^2+n_b^2},
\end{equation}

\begin{equation}
r_{\rm IQ} = \frac{n_a^3-n_b^3}{n_a^3+n_b^3}.
\end{equation}

\noindent For the example shown in the figure, the correlation coefficient happens to
equal the difference in mode frequency of occurrence, which is not true in general. 
From Equation~\ref{eqn:deltanu_a} when $n_b=1$, $\Delta\nu$ at point R is

\begin{equation}
\Delta\nu = 1 - 2\left(\frac{1}{1+n_a}\right)^{n_a}.
\end{equation}

Decreasing the value of $m$ further, the normalized mean of Stokes Q and the correlation 
coefficient are equal to one another at point E in the figure, where $C=1$ and $c=0$.

\begin{equation}
m = \bar{\rm Q} = r_{\rm IQ} = \frac{n_a-n_b}{n_a+n_b}
\end{equation}

\noindent The modulation index is always minimum at this location, where it is 
$\beta=1/\sqrt{n_a+n_b}$. From Equation~\ref{eqn:deltanu_a}, the difference in mode 
frequency of occurrence at point E is 

\begin{equation}
\Delta\nu = 1 - 2^{1-n_a}.
\end{equation}

% ----------------------------------------------------------------------------------

% increase space between data rows by 15%
\def\arraystretch{1.15}

\begin{deluxetable}{ccccccc}
\tablenum{2}
\tablecaption{Values of Select Statistical Parameters across a Mode Transition}
\tablehead{
  \colhead{Point} & \colhead{Description} &\colhead{$c$} & \colhead{$C$} 
  & \colhead{$m=\bar{\rm Q}$} & \colhead{$r_{\rm IQ}$} & \colhead{$\beta$}}
\startdata
A & Mode A only & 1 & $\infty$ & 1 & 1 & $\frac{1}{\sqrt{n_a}}$ \\
R & Mode A dominated & $\frac{n_a-n_b}{n_a+n_b}$ & $\frac{n_a}{n_b}$ 
  & $\frac{n_a^2-n_b^2}{n_a^2+n_b^2}$ & $\frac{n_a^3-n_b^3}{n_a^3+n_b^3}$
  & $\frac{\sqrt{n_a^3+n_b^3}}{n_a^2+n_b^2}$ \\
E & Minimum modulation & 0 & 1 & $\frac{n_a-n_b}{n_a+n_b}$ & $\frac{n_a-n_b}{n_a+n_b}$ 
  & $\frac{1}{\sqrt{n_a+n_b}}$ \\
G & No I-Q correlation & $\frac{\sqrt{n_b}-\sqrt{n_a}}{\sqrt{n_b}+\sqrt{n_a}}$ 
  & $\sqrt{\frac{n_b}{n_a}}$ & $\frac{\sqrt{n_a}-\sqrt{n_b}}{\sqrt{n_a}+\sqrt{n_b}}$ & 0 
  & $\frac{\sqrt{2}}{\sqrt{n_a}+\sqrt{n_b}}$ \\
T & Mode transition & $\frac{n_b-n_a}{n_b+n_a}$ & $\frac{n_b}{n_a}$ & 0 
  & $\frac{n_b-n_a}{n_b+n_a}$ & $\frac{1}{2}\sqrt{\frac{n_a+n_b}{n_an_b}}$ \\
B & Mode B only & -1 & 0 & -1 & -1 & $\frac{1}{\sqrt{n_b}}$ \\
\enddata
\end{deluxetable}

% restore arraystretch to default value of 1.0
\def\arraystretch{1.0}

% -----------------------------------------------------------------------------------

The correlation coefficient is equal to zero when the mode variances are equal to one 
another. This occurs at point G in the figure, where $C=\sqrt{n_b/n_a}$ and $c=-m$. 

\begin{equation}
c = -m = \frac{\sqrt{n_b}-\sqrt{n_a}}{\sqrt{n_b}+\sqrt{n_a}}
\end{equation}

\noindent The Stokes parameters I and Q are positively correlated for values of $m$ to the 
right of point G, and are negatively correlated for values of $m$ to the left of it. Mode A 
remains the primary mode, as indicated by $\Delta\nu>0$, and $\bar{\rm Q}=m$ remains positive 
despite the onset of the anticorrelation. The difference in mode frequency of occurrence at 
G is

\begin{equation}
\Delta\nu = 1 - 2\left(\frac{\sqrt{n_a}}{1+\sqrt{n_a}}\right)^{n_a}.
\end{equation}

The transition between modes occurs at point T in the figure, where the mode mean intensities 
are equal and $\bar{\rm Q}=m=0$. The value of the parameter $C$ at point T is $C=n_b/n_a$, and 
$c$ is equal to the I-Q correlation coefficient.
 
\begin{equation}
c = r_{\rm IQ} = \frac{n_b-n_a}{n_b+n_a}.
\end{equation}

\noindent The fractional linear polarization goes through a broad minimum between points G and 
T, but its minimum (at $m=0.033$ in the figure) does not coincide with T ($m=0$). From 
Equation~\ref{eqn:LbarA}, the mean fractional linear polarization at T is

\begin{equation}
\bar{\rm L}=\left(\frac{n_a}{1+n_a}\right)^{n_a}.
\end{equation}

\noindent The difference in frequency of occurrence at point T is

\begin{equation}
\Delta\nu = 1 - 2\left(\frac{n_a}{1+n_a}\right)^{n_a}.
\end{equation}

\noindent Since $\Delta\nu$ is generally not equal to zero at T, mode A unexpectedly remains 
the primary mode as $\bar{\rm Q}$ passes from positive to negative. 

The modes occur with equal frequency, and the primary mode changes from mode A to mode B, at 
point H in the figure, where $\Delta\nu=0$ (or $\nu_a=\nu_b=1/2$). When $n_b=1$, the change 
between primary modes occurs at 

\begin{equation}
c = 1 - 2^{\frac{n_a-1}{n_a}}.
\end{equation}

\noindent This value of $c$ causes $y=0$ to be the median of the Stokes Q distribution (see 
Equation~\ref{eqn:fQ}). Mode A is the primary mode for all values of $m$ to the right of point
H in the figure, and mode B is the primary for all values of $m$ to the left of it. The region
between points T and H has interesting implications for the polarization's average properties.
In this region of the figure, $\bar{\rm Q}$ is negative, but mode A is the primary mode, 
because $\Delta\nu >0$. Since the Stokes parameters Q and V are proportional to one another 
when the OPMs are elliptically polarized (see Equation~\ref{eqn:V}), $\bar{\rm Q}$ is a proxy 
for circular polarization, which in turn means the sense of the circular polarization between 
T and H on average is that of mode B (negative), although the PA is predominantly that of 
mode A.

The emission is comprised solely of mode B at point B, where $m=c=-1$. The emission is completely 
polarized and the Stokes parameters I and Q are completely anticorrelated. The modulation index 
is equal to that of mode B ($\beta=1/\sqrt{n_b}$). 

The values of $c$, $C$, $m$, $r_{\rm IQ}$ and $\beta$ at each of the points A, R, E, G, T, and 
B in the mode transition are summarized in Table 2. They are dependent only on the values 
of $n_a$ and $n_b$ and are independent of the scaling factors $a$ and $b$. The points R, E, G, T, 
and H merge at $m=c=0$ when $n_a=n_b$.

\begin{figure}
\plotone{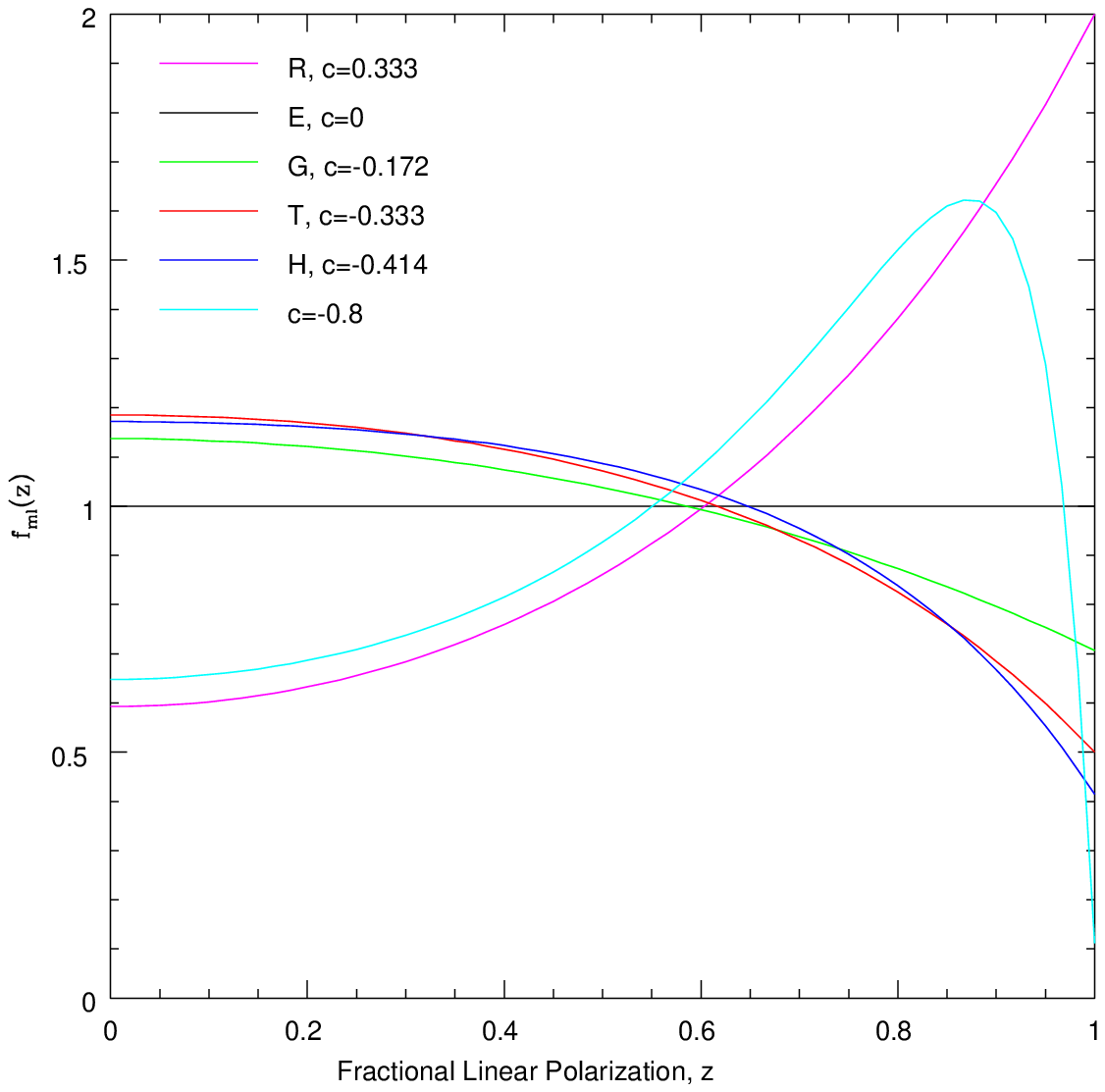}
\caption{The distributions of fractional linear polarization at different points within a mode 
transition. The points correspond to those labeled R, E, G, T, and H in Figure~\ref{fig:hysteresis} 
and Table 2. The distribution at $c=-0.8$ is also shown for comparison. The values of $n_a$, $n_b$, 
and $\chi_o$ used to produce the figure are $n_a=2$, $n_b=1$, and $\chi_o=0$.}
\label{fig:modefraclin}
\end{figure}

The distributions of fractional linear polarization at each of the points R, E, G, T, and H are 
shown in Figure~\ref{fig:modefraclin}. The distributions were calculated from Equation~\ref{eqn:fml} 
with $\chi_o=0$. The distribution at point R is shown by the magenta line in the figure. The 
slope of the distribution is generally positive for all values of $z$. Many data samples are highly, 
if not completely, polarized because the intensity of mode B, owing to its exponential distribution, 
is frequently very low, whereas the intensity of mode A is rarely so. The distribution at point E
in this specific case (the horizontal black line in the figure) is uniform, demarcating a change 
in the general slope of the distributions from positive to negative as $c$ continues to decrease. 
Despite the emission's random polarization implied by the distribution at this location, it is 
still dominated by mode A with a frequency of occurrence of $\nu_a=3/4$ (or $\Delta\nu=1/2$ from 
Equation~\ref{eqn:deltanu_a}). The distributions at points G, T, and H (the green, red, and blue 
lines, respectively) are similar in form with generally negative slopes over the full range of $z$. 
The intercepts of the distributions at $z=0$ initially increase with decreasing $c$, reaching a 
maximum at point T. Thereafter, the intercepts decrease as mode B becomes the primary mode. The 
distribution at $c=-0.8$ ($m=-0.636$, the cyan line) was arbitrarily selected to illustrate the 
distribution when mode B is the primary mode. The shape of the distribution there is quite different 
from when mode A is the primary. The distribution develops a peak as the value of $c$ decreases. 
Although many data samples in the distribution are highly polarized, they are not completely 
polarized, because the intensity of mode A is never equal to zero, unless it is completely absent 
(i.e. when $c=m=-1$). 

% --------------------------------------------------------------------------------------------

\subsection{Modulation and Depolarization}

\begin{figure}
\plotone{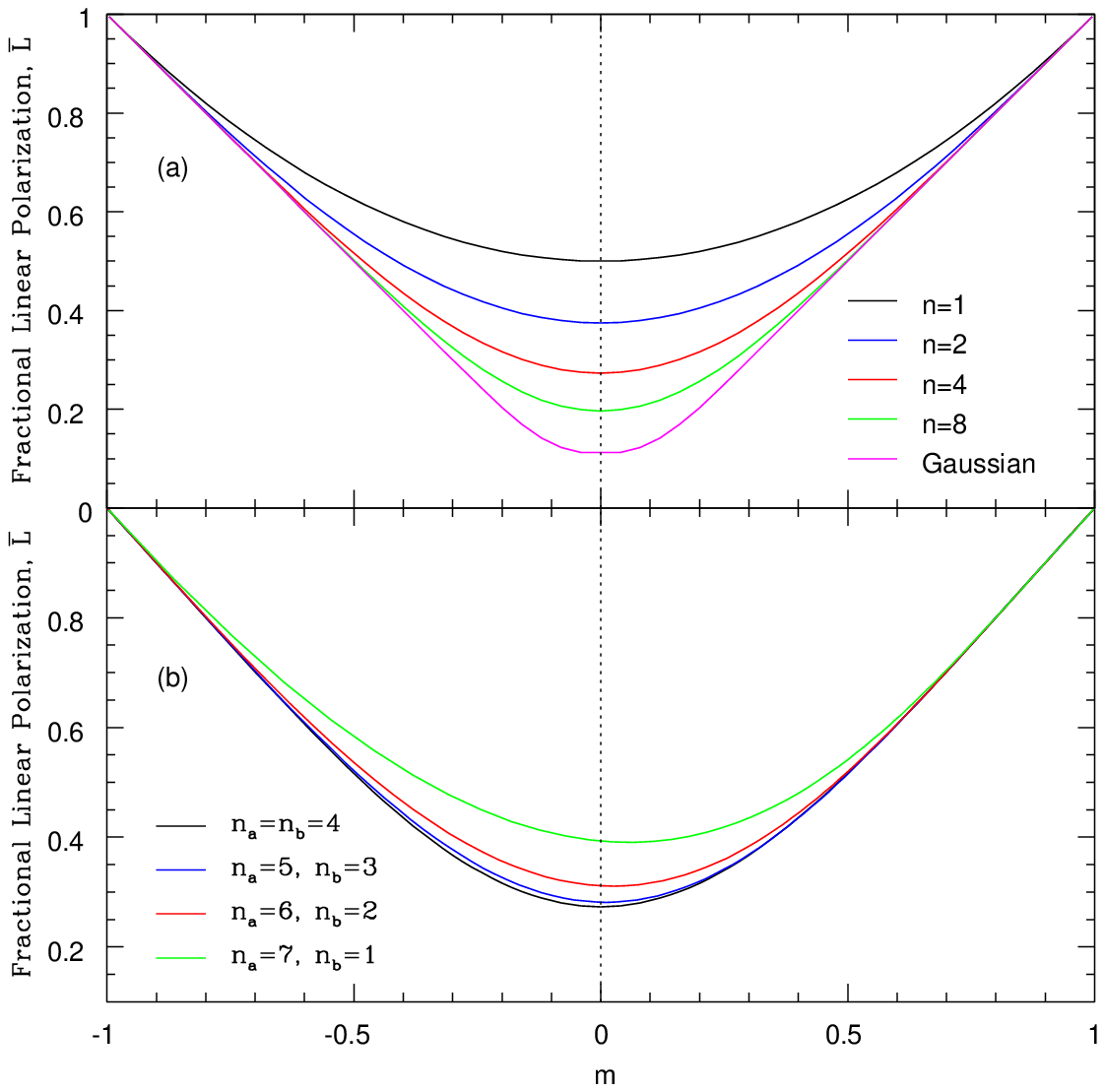}
\caption{Mean fractional linear polarization, $\bar{\rm L}$, for different values of $n_a$ and 
$n_b$. Panel (a) shows $\bar{\rm L}$ when $n_a=n_b=n$, for $n=1,2,4,$ and $8$ (Equation~\ref{eqn:Lbar}). 
The lower envelope in the panel is $\bar{\rm L}$ derived from Gaussian fluctuations in the mode 
intensities (Equation~\ref{eqn:LbarG}). Panel (b) shows $\bar{\rm L}$ when the sum of $n_a$ and 
$n_b$ remains constant at $n_a+n_b=8$.}
\label{fig:LbarN}
\end{figure}

\begin{figure}
\plotone{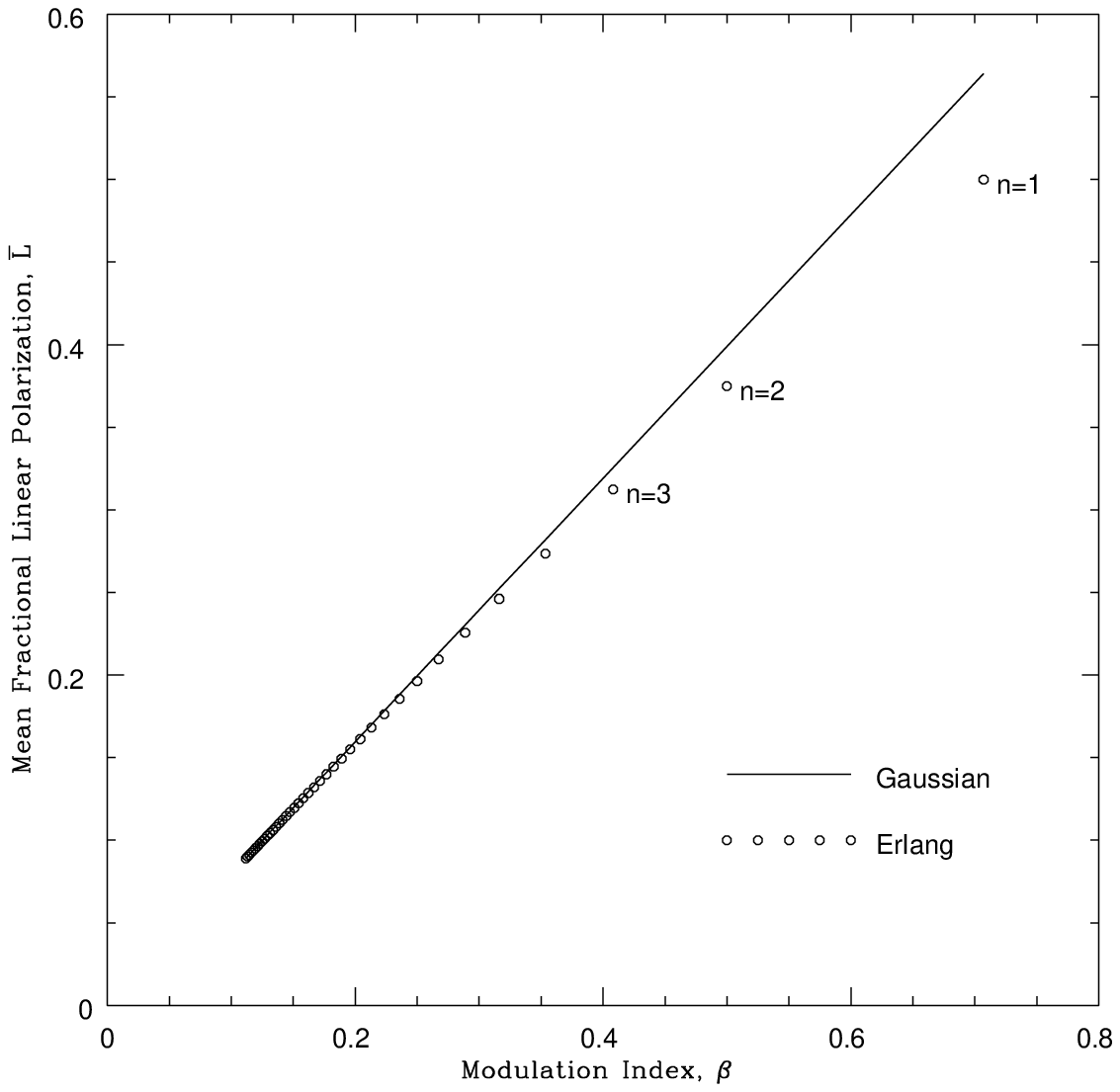}
\caption{Lower limit on the fractional linear polarization, $\bar{\rm L}$, that can be attained 
through superposed OPMs as a function of modulation index, $\beta$. The open circles denote the 
limit when the mode intensities are Erlang RVs with $1\le n\le 40$ (Equation~\ref{eqn:limit}). 
The solid line is the limit derived when the mode intensities are Gaussian RVs extrapolated to 
large values of $\beta$ (Equation~\ref{eqn:asymptote}).}
\label{fig:Lbeta}
\end{figure}

The effectiveness of superposed OPMs in depolarizing the emission and the relationship between
depolarization and intensity modulation can be explored with the equations for $\bar{\rm L}$ 
and $\beta$. Panel (b) of Figure~\ref{fig:LbarN} shows $\bar{\rm L}$ for different values of 
$n_a$ and $n_b$ with their sum held constant at $n_a+n_b=8$. The smallest value of $\bar{\rm L}$ 
(the most depolarization) occurs when $n_a=n_b=4$ at $m=0$. For each pair of $n_a$ and $n_b$ in 
the panel, the minimum value of the modulation index is always $\beta=1/\sqrt{8}$. As $n_a$ 
increases at the expense of $n_b$, the minimum of $\bar{\rm L}$ increases with respect to 
$\bar{\rm L}(0,4,4)$ at slightly increasing values of $m$. The conclusion to be drawn from 
panel (b) is, for a fixed value of $n_a+n_b$, the depolarization is maximum when the 
distributions of the mode intensities are identical (i.e. when $n_a=n_b$ and $a=b$). Panel (a)
of the figure shows $\bar{\rm L}$ when $n_a=n_b=n$ for different values of $n$. The magenta 
curve is $\bar{\rm L}$ derived for Gaussian mode intensities (see the discussion of 
Equation~\ref{eqn:LbarG} in Section~\ref{sec:gauss}). The curves are symmetric about $m=0$ 
where $\bar{\rm L}$ is minimum. From Equation~\ref{eqn:Lbar}, the value of $\bar{\rm L}$ at 
$m=c=0$ is

\begin{equation}
\bar{\rm L}(0,n,n)=\frac{1}{n2^{2n-1}}\frac{(2n-1)!}{(n-1)!^2},
\label{eqn:limit}
\end{equation}

\noindent and the modulation index is $\beta=1/\sqrt{2n}$. The additional conclusion to be drawn 
from panel (a) of the figure is the mean fractional linear polarization is minimum when the 
modulation index is small ($n$ is large); the less the mode intensities fluctuate, the more
depolarization will occur. Equation~\ref{eqn:limit} then represents a lower limit on the fractional 
linear polarization that can be attained through superposed OPMs as a function of $n$. The limit 
on $\bar{\rm L}$ is related to modulation index through $n$. The limit as a function of $\beta$ 
is shown by the open circles in Figure~\ref{fig:Lbeta} for $1\le n\le 40$. The solid line in the 
figure is the same limit derived from Gaussian fluctuations in mode intensities, and is discussed 
in Section~\ref{sec:gauss} (see Equation~\ref{eqn:asymptote}). The figure shows the fractional 
linear polarization remains high ($\bar{\rm L}=1/2$) when the modulation index is large 
($\beta=1/\sqrt{2}$). 

\subsection{Statistical Parameters and Distributions for Gaussian Mode Intensities}
\label{sec:gauss}

As the order of the Erlang distributions of mode intensities becomes very large, the various 
distributions and statistical parameters derived from the statistical model should resemble those 
derived from Gaussian-distributed mode intensities. The distributions and parameters arising from 
Gaussian fluctuations were derived in MS; however, they were not presented as functions of $m$ in 
that original implementation of the model. The distributions and parameters written in terms of 
$m$ are summarized here to facilitate their comparison with what was derived from Erlang mode 
intensities.

When the mode intensities are Gaussian RVs, the normalized Stokes parameter Q is $\bar{\rm Q}=m$, 
and the I-Q correlation coefficient is always zero, $r_{\rm IQ}=0$ (MS). The modulation index is 
constrained to be

\begin{equation}
\beta(m)\le \frac{1-\lvert m\rvert}{5\sqrt{2}}.
\label{eqn:BetaGauss}
\end{equation}

\noindent The constraint on $\beta$ arises from how the statistical model was implemented in 
MS. The standard deviations of the mode intensities, $\sigma$, were held constant and equal 
to make the Stokes parameters I and Q independent of one another. The parameter $m$ (or $M$) 
was then varied by effectively holding the mean intensity of the secondary mode constant and 
varying the mean of the primary mode. The mean intensity of the secondary mode must have a 
minimum value of about $5\sigma$ to ensure its intensity is always nonnegative. 

The frequency of occurrence of mode A for Gaussian mode intensities is given by Equation 15 
of MS

\begin{equation}
\nu_a = \frac{1}{2}\left[1+{\rm erf}\left(\frac{\mu_Q}{\sigma_Q\sqrt{2}}\right)\right],
\label{eqn:MSeqn15}
\end{equation}

\noindent where $\rm{erf}(x)$ is the error function, $\mu_{\rm Q}$ is the mean of the 
Stokes parameter Q, and $\sigma_{\rm Q}$ is its standard deviation. As with 
Equation~\ref{eqn:BetaGauss}, Equation~\ref{eqn:MSeqn15} assumes the standard deviations of 
the mode intensities are equal, which in turn means the standard deviations of the Stokes 
parameters I and Q are also equal, $\sigma_{\rm Q}=\sigma_{\rm I}=\sigma\sqrt{2}$. Recognizing 
that $\mu_{\rm Q}=m\mu_{\rm I}$ and $\beta=\sigma_{\rm I}/\mu_{\rm I}$, the equation can be 
rewritten in terms of $m$ and $\beta$ as

\begin{equation}
\nu_a(m,\beta) = \frac{1}{2}\left[1+{\rm erf}\left(\frac{m}{\beta\sqrt{2}}\right)\right],
\label{eqn:nuGauss}
\end{equation}

\noindent The difference in mode frequency of occurrence is

\begin{equation}
\Delta\nu(m,\beta) = \rm{erf}\left(\frac{m}{\beta\sqrt{2}}\right).
\label{eqn:deltanuG}
\end{equation}

\begin{figure}
\plotone{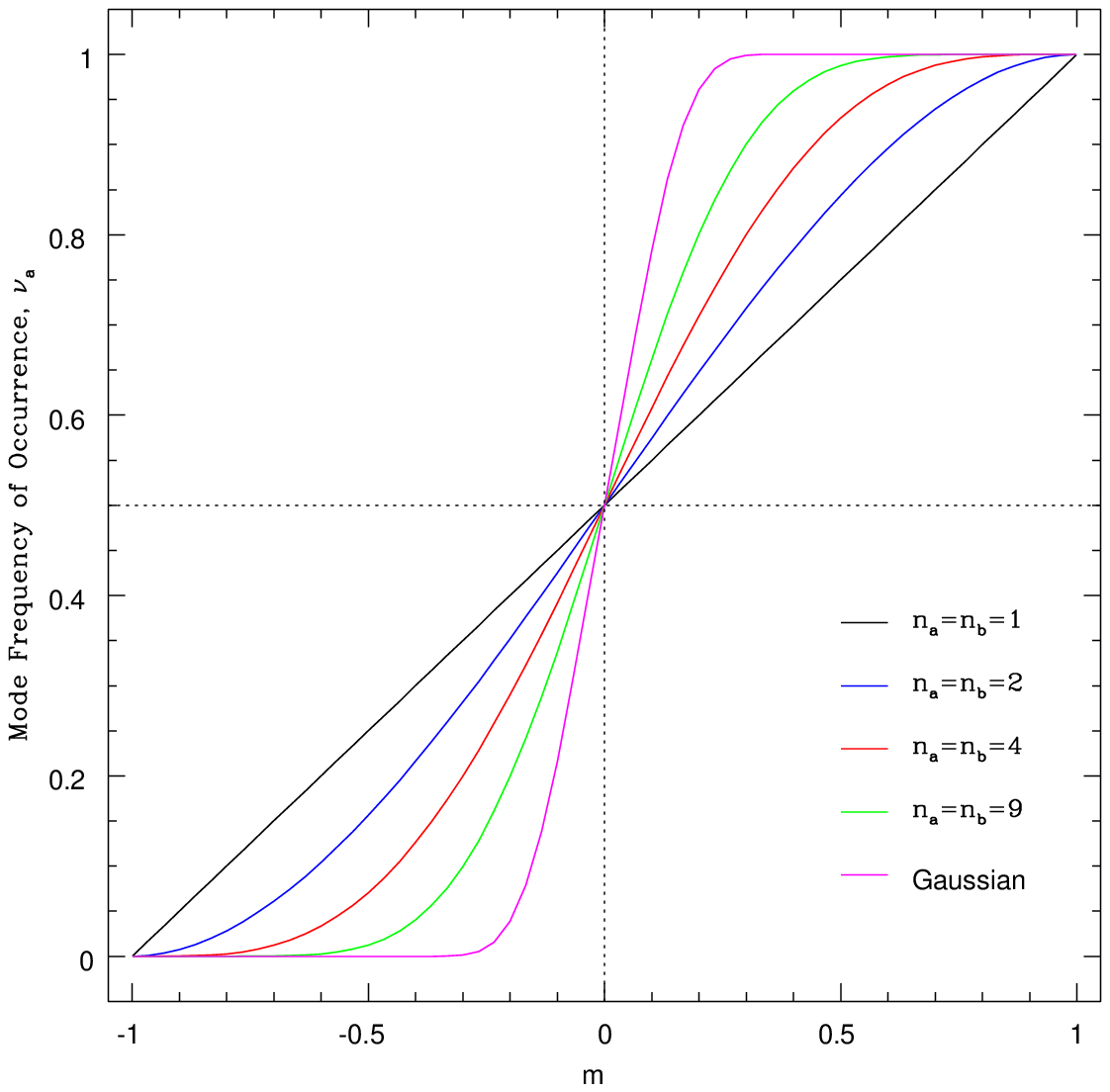}
\caption{Behavior of the mode frequency of occurrence, $\nu_a$, at a mode transition when the 
mode intensities are Erlang (Equation~\ref{eqn:nuA}) and Gaussian (Equation~\ref{eqn:nuGauss})
RVs. The transition occurs at $m=0$. The values of $n$ used for the Erlang distributions are 
annotated in the figure.}
\label{fig:nuA}
\end{figure}

\noindent The frequencies of occurrence for mode A derived from Gaussian and Erlang fluctuations 
in mode intensities are compared in Figure~\ref{fig:nuA}. The dependence of $\nu_a$ upon $m$ for 
Gaussian fluctuations was calculated from Equation~\ref{eqn:nuGauss} using 
Equation~\ref{eqn:BetaGauss} for the functional dependence of $\beta$ upon $m$. 
Equation~\ref{eqn:nuA} was used to calculate $\nu_a$ for Erlang fluctuations when $n_a=n_b=n$. 
The different values of $n$ used to calculate $\nu_a$ are annotated in the figure. The figure 
shows that $\nu_a$ for Erlang fluctuations approaches that for Gaussian fluctuations as the value 
of $n$ increases, as expected from the central limit theorem. Each example of $\nu_a$ in 
the figure is antisymmetric about $m=0$, because the mode intensity distributions are identical 
to one another in each case. The frequency of occurrence changes rapidly near $m=0$ for Gaussian 
intensity fluctuations, but changes more gradually over $m$ for Erlang fluctuations. When $n=1$, 
the mode frequency of occurrence varies linearly with $m$, $\nu_a=(1+m)/2$ (M22). When $n$ is 
large, a small change in $m$ near $m=0$ can result in a large change in $\nu_a$. Since the 
modulation index of an Erlang distribution is generally larger than that for a Gaussian 
distribution, one may infer that rapid changes in $\nu_a$ occur where the modulation index is 
small, and more gradual changes occur where the modulation index is large. 

The mean of the linear polarization for Gaussian fluctuations in mode intensities is given by 
Equation 13 of MS.

\begin{equation}
\mu_{\rm L} = \sigma_{\rm Q}\sqrt{\frac{2}{\pi}}\exp{\left(-\frac{\mu_{\rm Q}^2}{2\sigma_{\rm Q}^2}\right)}
        + {\rm erf}\left(\frac{\mu_{\rm Q}}{\sigma_{\rm Q}\sqrt{2}}\right)\mu_{\rm Q}.
\end{equation}

\noindent Dividing $\mu_{\rm L}$ by the mean total intensity and rewriting the equation 
in terms of $m$ and $\beta$ gives the normalized mean of the linear polarization.

\begin{equation}
\bar{\rm L}(m,\beta) = \beta\sqrt{{2\over{\pi}}}\exp{\left(-{m^2\over{2\beta^2}}\right)}
        + {\rm erf}\left({m\over{\beta\sqrt{2}}}\right)m.
\label{eqn:LbarG}
\end{equation}

\noindent Equation~\ref{eqn:LbarG} with $\beta$ given by Equation~\ref{eqn:BetaGauss} is 
shown by the magenta curve in panel (a) of Figure~\ref{fig:LbarN}. The first term in the 
equation is the contribution of the polarization fluctuations to $\bar{\rm L}$, and the 
second term is the contribution of the persistent polarization to $\bar{\rm L}$. The 
equation shows that $\bar{\rm L}$ can never be zero when the polarization fluctuates (i.e. 
when $\beta\ne 0$). Since $m=0$ at an OPM transition for Gaussian mode intensities, the 
fractional linear polarization at the transition is directly proportional to the modulation 
index, $\bar{\rm L}=\beta\sqrt{2/\pi}$, and is small where $\beta$ is also small. This is 
the smallest the fractional linear polarization can be when the mode intensities are Gaussian 
RVs. The asymptotic behavior of $\bar{\rm L}$ at large values of $n$ for Erlang mode 
intensities can be determined by setting the limit given by Equation~\ref{eqn:limit} equal 
to $\bar{\rm L}$ calculated for Gaussian mode intensities. Since the modulation index for 
Erlang mode intensities is $\beta=1/\sqrt{2n}$ when $m=0$, the limit on $\bar{\rm L}$ at 
large values of $n$ is 

\begin{equation}
\bar{\rm L} = \beta\left(\frac{2}{\pi}\right)^{1/2}=\left(\frac{1}{n\pi}\right)^{1/2}.
\label{eqn:asymptote}
\end{equation}

\noindent Equation~\ref{eqn:asymptote} is shown by the solid line in Figure~\ref{fig:Lbeta}.
The figure shows the minimum in fractional linear polarization derived from Gaussian 
fluctuations in mode intensities is in good agreement with the result obtained from Erlang 
fluctuations when $n$ is large. The minimum for true Gaussian fluctuations is applicable only 
for $\beta\le 1/(5\sqrt{2})$ (or $n\ge 25$). The line has been extended in the figure to 
include $n=1$ to provide a full comparison between the two results. 

\begin{figure}
\plotone{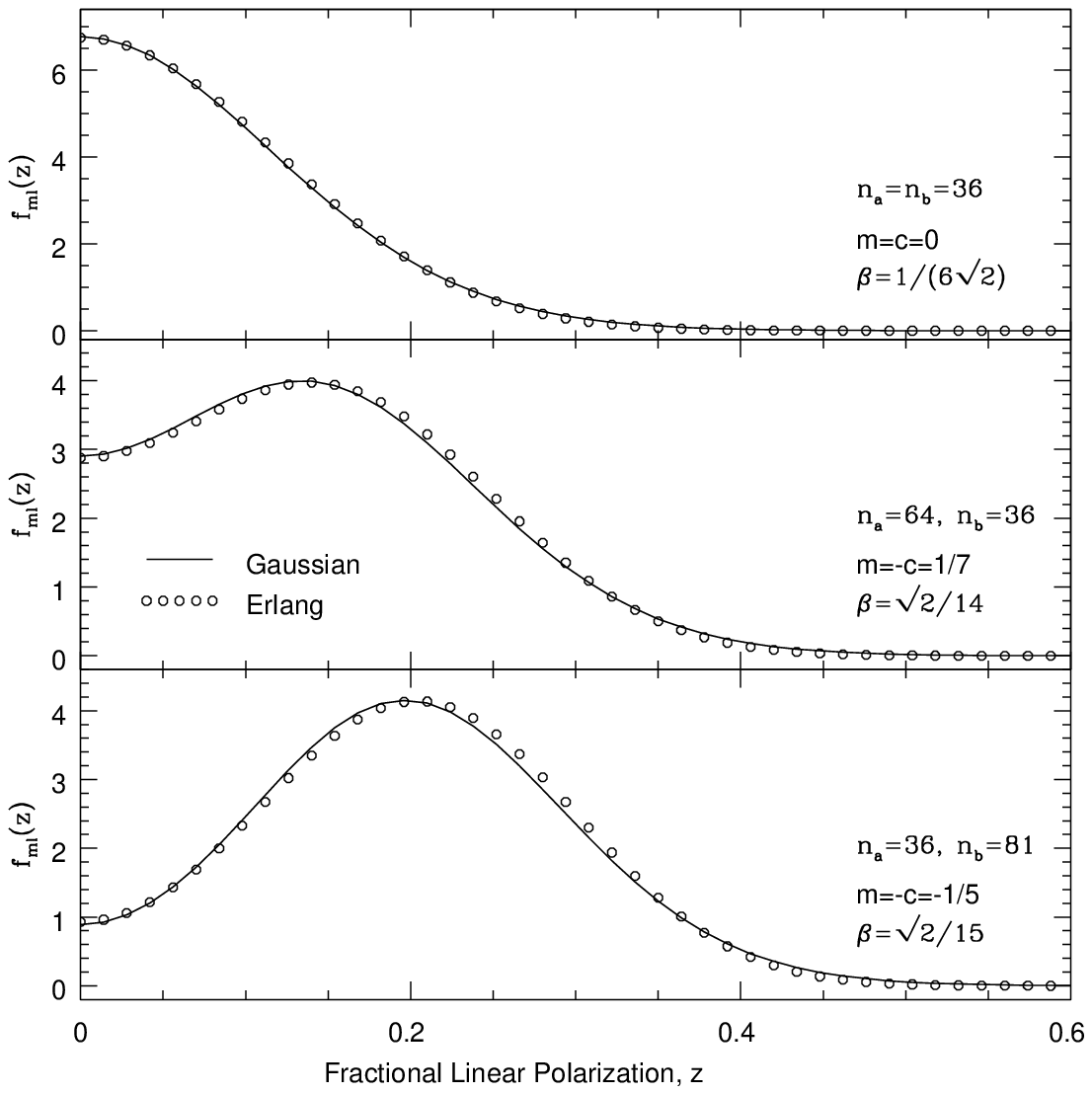}
\caption{Comparison of distributions of fractional linear polarization when the fluctuations
in mode intensities are Gaussian and Erlang RVs. The distributions resulting from Gaussian
fluctuations (Equation~\ref{eqn:LinGauss}) are shown by the solid lines, and the distributions
resulting from Erlang fluctuations (Equation~\ref{eqn:fml}) are denoted by open circles. The
values of $n_a, n_b, c$, and $\beta$ used to determine the distributions are annotated
in each panel of the figure. The value of the ellipticity angle used in the figure is 
$\chi_o=0$.}
\label{fig:EGLcmp}
\end{figure}

When the mode intensities are Gaussian RVs, the distributions of the Stokes parameters are 
also Gaussian (MS). The distribution of linear polarization is the sum of two Gaussians 
with equal variances. Their means are also equal in magnitude, but with opposite signs (see 
Equation 27 of MS). The distribution of fractional linear polarization is given by Equation 
16 of MS. Reparameterizing that equation in terms of $m$ and $\beta$, and accounting for 
the ellipticity of the modes' polarization, gives

\begin{eqnarray}
fml(z,m,\beta,\chi_o) & = & 
                      {1\over{\beta\sqrt{2\pi}}}{\cos(2\chi_o)\over{(z^2+\cos^2(2\chi_o))^{3/2}}}
                      \Biggl\{(\cos(2\chi_o)+mz)\exp{\left[-{(z-m\cos(2\chi_o))^2
                      \over{2\beta^2(z^2+\cos^2(2\chi_o))}}\right]} \nonumber \\
                      &  + & (\cos(2\chi_o)-mz)\exp{\left[-{(z+m\cos(2\chi_o))^2
                      \over{2\beta^2(z^2+\cos^2(2\chi_o))}}\right]}\Biggr\}.
\label{eqn:LinGauss}
\end{eqnarray}

\noindent The distribution of fractional circular polarization is

\begin{equation}
fmv(z,m,\beta,\chi_o) = {\sin(2\chi_o)\over{\beta\sqrt{2\pi}}}{\sin(2\chi_o) + mz
                 \over{(z^2+\sin^2(2\chi_o))^{3/2}}}
                 \exp{\left[-{(z-m\sin(2\chi_o))^2
                 \over{2\beta^2(z^2+\sin^2(2\chi_o))}}\right]}.
\label{eqn:CircGauss}
\end{equation}

\begin{figure}
\plotone{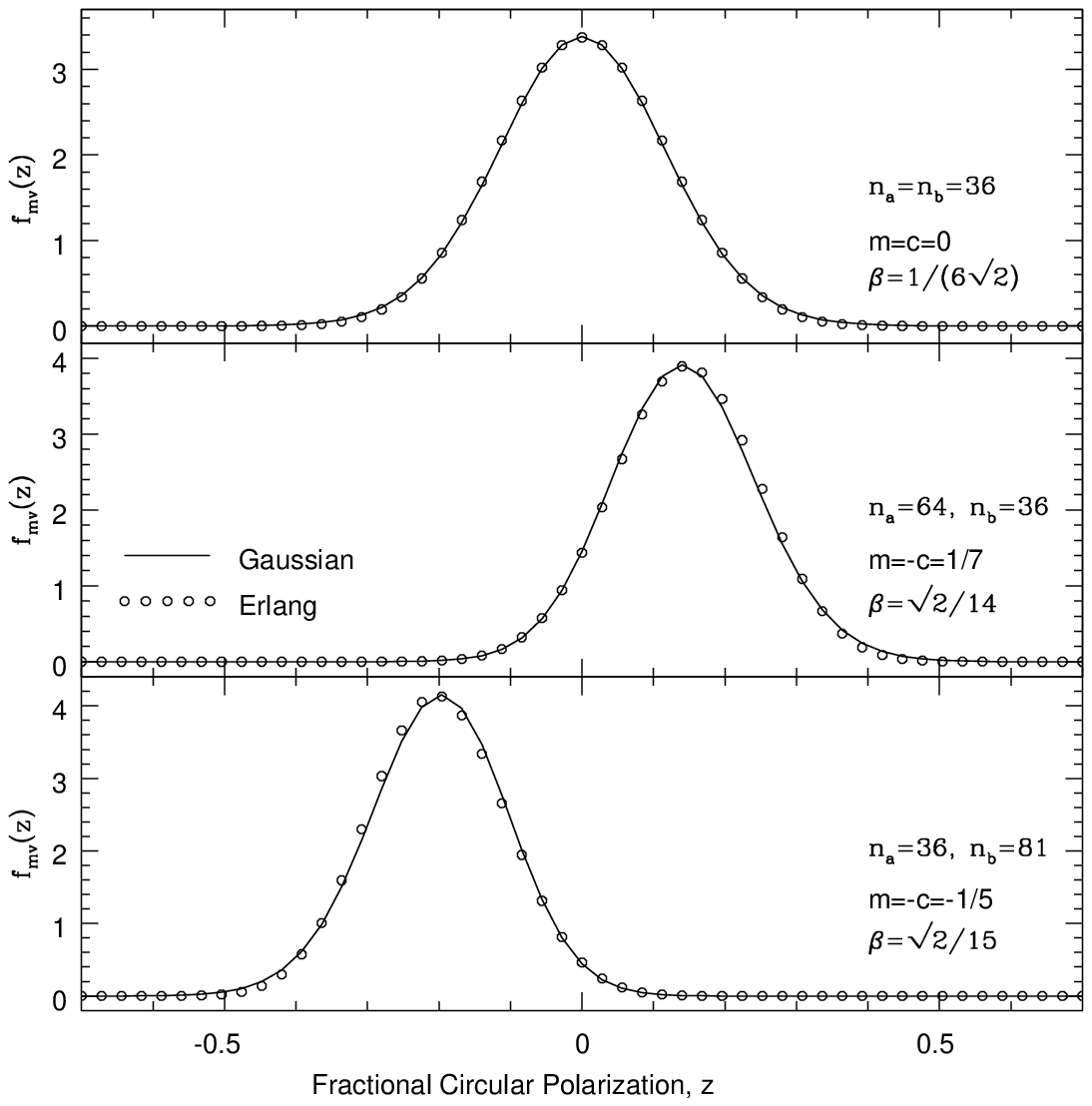}
\caption{Comparison of distributions of fractional circular polarization when the fluctuations in 
mode intensities are Gaussian and Erlang RVs. The distributions resulting from Gaussian fluctuations 
(Equation~\ref{eqn:CircGauss}) are shown by the solid lines, and the distributions resulting from 
Erlang fluctuations (Equation~\ref{eqn:fmv}) are denoted by open circles. The value of the 
ellipticity angle used in the figure is $\chi_o=\pi/4$.}
\label{fig:EGVcmp}
\end{figure}

\noindent The distribution is approximately Gaussian with a mean of $m\sin(2\chi_o)$ and a 
standard deviation of $\beta\sin(2\chi_o)$. The distributions of fractional polarization for 
Gaussian and Erlang fluctuations in mode intensities are compared in Figures~\ref{fig:EGLcmp} 
and~\ref{fig:EGVcmp} when the conditions used to derive the equations for them are satisfied. 
This occurs when the Erlang mode intensities are identically distributed ($n_a=n_b\gg 1$ and 
$m=0$) or when their variances are equal (i.e. when $r_{\rm IQ}=0$), as at point G in 
Figure~\ref{fig:hysteresis} and Table 2. The modulation index of the total intensity resulting 
from Gaussian fluctuations must also satisfy $\beta\le 1/(5\sqrt{2})$. The parameters used to 
calculate the distributions are annotated in each panel of the figures. The distributions
resemble one another, which is noteworthy considering the substantial differences between their 
functional forms.

Equations~\ref{eqn:nuGauss}, ~\ref{eqn:deltanuG}, ~\ref{eqn:LbarG}, ~\ref{eqn:LinGauss}, 
and~\ref{eqn:CircGauss} are written as functions of $m$ and $\beta$ for conciseness and to 
indicate how the statistical parameters and fractional polarization distributions are influenced 
by them, but with the understanding that $m$ and $\beta$ are not independent and follow the 
constraint stipulated by Equation~\ref{eqn:BetaGauss}.

% --------------------------------------------------------------------------------------------

\section{DISCUSSION}
\label{sec:discuss}

\subsection{Summary Comments on the Analysis}

The MS statistical model recognizes and incorporates the stochastic nature of pulsar radio 
emission and the tendency of its polarization to randomly switch between orthogonally polarized 
states. The model can replicate observed distributions of the Stokes parameters, linear 
polarization, and polarization position angle at a given pulse phase (MS, M22). It predicts 
polarization fluctuations have a preferred orientation in the Poincar\'e sphere (M04). The 
orientation becomes more apparent as the modulation index increases. The implementation of the 
model in this analysis, where the mode intensities are assumed to be Erlang RVs, is the 
general version of the specific cases considered in MS and M22. The analysis in M22 represents 
one extreme of the general case, where the mode intensity fluctuations are modeled as exponential 
RVs ($n=1$). The analysis in MS is the other extreme, where the fluctuations are modeled as 
Gaussian RVs ($n\gg 1$). All intervening values of the order of the Erlang intensity distributions, 
$n$, can be accommodated by the analysis presented here. 

A pulse phase-resolved transition between the polarization modes in an average profile can be 
modeled by virtue of the analysis. The modeling is enabled by the realization that the ratio of 
the mode mean intensities, as represented by the parameter $m$, drives the transition. Since $m$ 
must vary with pulse phase, $\phi$, to complete the transition, $m$ is a proxy for $\phi$ over 
the transition's duration. The simplest relationship between $m$ and $\phi$ is they are 
proportional to one another (see, e.g., McKinnon 2003). 
 
The analysis allows an investigation into how mode intensities of different statistical 
character might affect the observed polarization. When the mode intensities are identically 
distributed, the mean fractional linear polarization and modulation index are symmetric 
about a mode transition in an average profile. The symmetry is disrupted when the mode 
intensity distributions are different. The latter scenario may manifest itself in seemingly 
anomalous ways, such as a minimum in the fractional linear polarization that does not coincide 
with the mode transition, or the prevailing sense of the circular polarization in an average 
profile is that of one mode although the PA histogram at the same pulse phase indicates the 
other mode is occurring more frequently.

The analysis also posits that the fluctuations in mode intensities at least contribute 
to, or may completely determine, the overall modulation of the emission. The statistical 
character of the mode intensity fluctuations is presumed to be determined by the mechanism 
responsible for their excitation, either via different emission mechanisms or mode-dependent 
propagation or scattering effects. The observed variations in modulation index across a mode 
transition would then arise from changes in the relative excitations of the two modes. Changes 
in $\beta$ should be reflected by changes in the axial ratio of Q-U-V data point clusters or by 
changes in a mode's frequency of occurrence. If observed variations in modulation index across a 
mode transition are large, then the mode intensity distributions must be different to explain 
the observed range of $\beta$ values. The order of the Erlang distribution for one mode would 
need to be $n=1$ to account for modulation indices as large as $\beta=1$, and the order of the 
distribution for the other mode would need to be much greater than one to account for small 
values of $\beta$. 

The assumption of Erlang mode intensities in the analysis should not be misconstrued as 
a claim that the intensities always follow Erlang statistics. Other probability distributions 
of intensity and pulse energy, such as log-normal, power law, and gamma, are possible and have 
been observed (e.g., Cairns et al. 2003a, 2003b; Burke-Spolaor et al. 2012). A general conclusion 
from the analysis, that the mean fractional linear polarization and modulation index are 
symmetric about a mode transition when the mode intensity distributions are identical but are
asymmetric when the distributions are different, likely holds for these other distributions, 
as well. The log-normal distribution is of particular interest because it is frequently observed 
and can produce values of modulation index that are larger than those considered here.

The limit on the fractional linear polarization attributable to OPMs for low values of 
modulation index (Equation~\ref{eqn:asymptote}) is a familiar result. If the OPMs have the same 
fixed intensity, the resulting polarization and fractional polarization are both equal to zero. 
If the intensities then randomly vary while retaining the same mean intensity, the mean
polarization and mean fractional polarization will both increase because some intensity samples 
will be polarized and others will not. Therefore, the mean polarization and mean fractional 
polarization increase as the fluctuations increase. When the intensities are identically 
distributed Gaussian RVs, the mean linear polarization is given by the well-known result 
$L=\sigma_o\sqrt{\pi/2}$ (e.g. Papoulis 1965, p. 195), where $\sigma_o$ is the standard 
deviation of the total intensity, such that the mean fractional linear polarization is 
$\bar{\rm L}=\beta\sqrt{\pi/2}$. The constant of proportionality between $\bar{\rm L}$ and 
$\beta$ here is different from what was derived in Equation~\ref{eqn:asymptote}, because the 
idealized derivation in Section~\ref{sec:implement} and MS concentrated the polarization in the 
Stokes parameter Q and held the Stokes parameter U fixed at zero. The limit on $\bar{\rm L}$ 
for Gaussian mode intensities cannot increase indefinitely with increasing $\beta$, because the 
extent of the Gaussian fluctuations cannot exceed the mean to ensure the individual intensity 
samples are nonnegative. The upper limit on $\beta$ for Gaussian fluctuations in this analysis 
is $\beta\le 1/(5\sqrt{2})$ (Equation~\ref{eqn:BetaGauss}). The limit on $\bar{\rm L}$ can be 
extended to larger values of $\beta$ when the mode intensities are Erlang RVs, which are always 
nonnegative by definition, and consequently are not similarly encumbered by the limit on $\beta$. 
Erlang mode intensities extend the range of the modulation index to $\beta\le 1/\sqrt{2}$, which 
produces a corresponding limit on the fractional linear polarization of $\bar{\rm L}=1/2$. When 
$\beta$ is small ($n\gg 1)$, the $\beta$ dependence of $\bar{\rm L}$ produced by Erlang mode 
intensities replicates the behavior derived from Gaussian intensities. 

% -------------------------------------------------------------------------------------------

\subsection{Examples from Observations}

The analysis evaluated the effectiveness of fluctuating OPM intensities in depolarizing the emission,
and determined the depolarization is greatest when the mode intensities are identically distributed 
and the modulation index is small (Figures~\ref{fig:LbarN} and~\ref{fig:Lbeta}). This result suggests
that mode transitions accompanied by acute depressions in linear polarization occur where the mode 
intensities are equal and somewhat stable, almost as if the OPMs behaved as fixed radiators. An 
example of this behavior occurs at the leading edge of PSR B0626+24 (Figure 2 of Weisberg et al. 
1999). The pulse is almost completely linearly polarized at this location, but the polarization 
plunges to near-zero levels at a following mode transition. The modulation index is at its minimum 
value near the transition (Figure A.2 of Weltevrede et al. 2006). 

The mode frequency of occurrence changes at a mode transition, and particularly so when the mode 
intensities are identically distributed and the modulation index is low (Figure~\ref{fig:nuA}). 
An example of this general behavior occurs in the leading component of PSR B2020+28 (Figure 3 of 
MS), where the linear polarization decreases abruptly at a mode transition. The PA histograms at 
this location (Figure 5 of MS) evolve from both modes occurring with nearly equal frequency at 
the transition to only one mode occurring within about six pulse phase bins. The modulation 
indices at this location ($0.25\le\beta\le 0.46$) approach the lowest value within the pulse 
($\beta=0.18$; Figure 2(c) of M04). These examples suggest mode intensity fluctuations contribute 
to both the depolarization and the modulation of the emission.

% --------------------------------------------------------------------------------------------

\section{CONCLUSIONS}
\label{sec:conclude}

A statistical model was used to explore the polarization and modulation properties of pulsar
radio emission by assuming the intensities of the OPMs follow different Erlang distributions.
General expressions were derived for the distributions of the resulting Stokes parameters and
linear polarization. All are additive combinations of weighted Erlang distributions. General
expressions for the distributions of fractional polarization and polarization position angle
were also derived, and were shown to be consistent with their Gaussian counterparts when the
order of the Erlang distributions is large.

A transition between the modes was examined in detail. The changing ratio of the mode mean 
intensities drives the transition from one mode to the other. The parameter $m$ serves as 
a proxy for pulse phase and allows the behavior of the polarization and modulation statistical 
parameters to be resolved over the transition's duration. The fractional linear polarization 
and modulation index are symmetric about a transition when the mode intensity distributions 
are the same. The symmetry is disrupted when the distributions of mode intensities are 
different. 

The modulation and polarization of pulsar radio emission are affected by both systematic 
and stochastic processes. The analysis presented in this paper focuses on a stochastic 
component to the emission that is affiliated with OPMs. The analysis, coupled with 
observational evidence, suggests the mode fluctuations contribute to the depolarization 
and overall modulation of the emission. If the emission is comprised solely of the 
stochastic component, the distributions of the mode intensities would need to be different 
to explain the larger variations in modulation index observed across a mode transition.

The effectiveness of superposed OPMs in depolarizing the emission was investigated. The 
fractional linear polarization is minimum at mode transitions where the OPM intensities 
are identically distributed and the modulation index of the total intensity is small. A 
limit on the minimum fractional linear polarization that can be attributed to the 
superposition of the modes as a function of modulation index was quantified. For those 
pulsars showing acute, deep depressions in linear polarization at a transition, the 
modulation index is likely small and the OPM emission is consequently quasi-stable at 
those locations. 

% --------------------------------------------------------------------------------------------

\acknowledgments{The National Radio Astronomy Observatory is a facility of the National 
Science Foundation operated under cooperative by Associated Universities, Inc.}

% ---------------------------------------------------------------------------------

\end{document}